\documentclass[11pt,draftcls,onecolumn]{IEEEtran}

\usepackage{algorithm}
\usepackage{algorithmic}
\usepackage{cite}
\usepackage{amsmath} 
\usepackage{amssymb} 
\usepackage{dsfont}
\usepackage[mathcal]{euscript}
\usepackage{empheq}
\usepackage{graphicx}
\usepackage{graphics}
\usepackage{bm}
\usepackage{psfrag}
\usepackage{mathrsfs}
\usepackage{bbm}
\usepackage{color}
\usepackage{cancel}
\input{epsf}

\newtheorem{lemma}{Lemma}

\newtheorem{proposition}{Proposition}
\newtheorem{remark}{Remark}

\newtheorem{assumption}{Assumption}

\renewcommand{\P}{\mathbb{P}}
\newcommand{\E}{\mathbb{E}}

\newcommand{\beq}{\begin{equation}}
\newcommand{\eeq}{\end{equation}}
\newcommand{\beqa}{\begin{eqnarray}}
\newcommand{\eeqa}{\end{eqnarray}}
\newcommand{\dfz}{\triangleq}

\DeclareMathOperator*{\argmin}{argmin}

\begin{document}

\title{Topology Inference over Networks with Nonlinear Coupling}

\author{Augusto Santos$^{\star}$ , Vincenzo Matta$^\dagger$, and Ali H. Sayed$^{\star}$

\thanks{$^\star$ A. Santos (email: augusto.santos@epfl.ch)
and A.~H.~Sayed (email: ali.sayed@epfl.ch) are with the \'Ecole Polytechnique F\'ed\'erale de Lausanne (EPFL), CH-1015 Lausanne, Switzerland. The work of A.~H.~Sayed was also supported in part by US NSF grant CCF-1524250 and Swiss NSF (SNSF) grant 205121\textunderscore184999.}
\thanks{$\dagger$ V. Matta is with DIEM, University of Salerno, via Giovanni Paolo II, I-84084, Fisciano (SA), Italy (email: vmatta@unisa.it).}
}

\maketitle

\begin{abstract}
This work examines the problem of topology inference over discrete-time nonlinear stochastic networked dynamical systems. The goal is to recover the underlying {\em digraph} linking the network agents, from observations of their state-evolution. 
The dynamical law governing the state-evolution of the interacting agents might be nonlinear, i.e., the next state of an agent can depend nonlinearly on its current state and on the states of its immediate neighbors. 
We establish sufficient conditions that allow consistent graph learning over a special class of networked systems, namely, logistic-type dynamical systems. 
\end{abstract}


\section{Introduction}

\begin{figure}[t]
\centering
\includegraphics[width=80mm,angle=270]{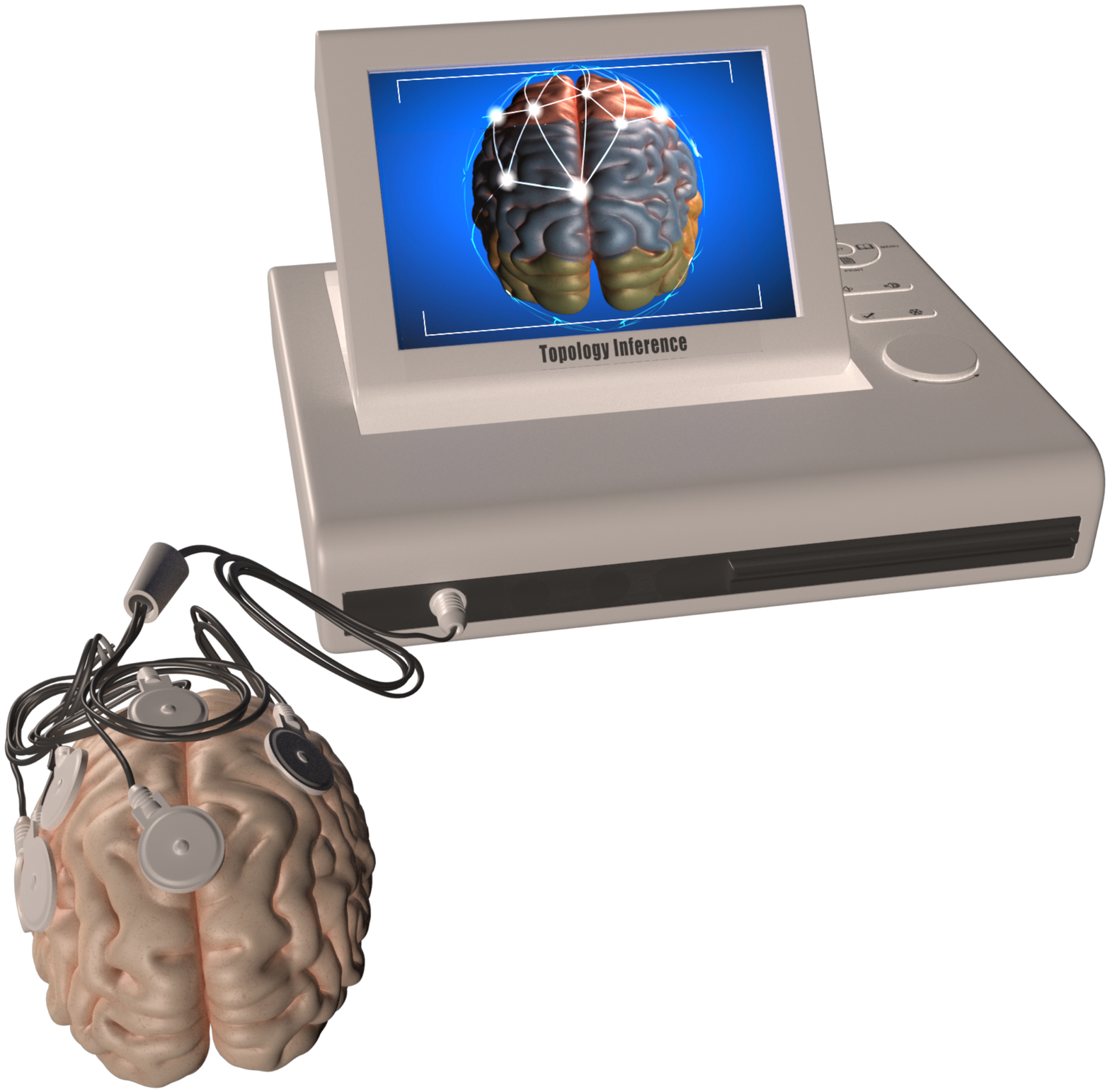}
\caption{Illustration of the topology inference paradigm: data are collected from a set of nodes in a networked dynamical system --- e.g., signals measured at regions of interest in the brain --- and the underlying connectivity among the nodes in the system --- or functional connectivity among the regions in the brain --- is then inferred.} 
\label{fig:brain}
\end{figure}

\IEEEPARstart{T} \,he evolution of a {\em networked} dynamical system is determined by the local interactions among its neighboring agents. Graph or structure learning refers to the problem of estimating the underlying graph from observations collected at the agents.
This is a challenging inverse problem, which would allow us to understand more fully the evolution of systems arising across several application domains including, e.g., epidemics~\cite{Barrat}, social networks~\cite{SalamiYingSayedTSIPN2017}, and brain activity~\cite{Honeyetal2009}. 
Figure~\ref{fig:brain} provides an illustration of the topology inference problem with reference to the last application.

Very often, the dynamics of real-world phenomena is governed by {\em highly nonlinear} and possibly {\em random} forms of interaction. For this reason, one useful class of graph learning problems concerns the case of nonlinear stochastic dynamical systems. 
In this article, we focus on the following class of logistic-type systems, which can model several forms of {\em nonlinear coupling} between locally-interacting (i.e., neighboring) agents: 
\beq
\boxed{
\bm{y}_{i,n+1}=
\sigma_i\left(
g_i(\bm{y}_{i,n}) \sum_{j = 1}^N a_{ij} h_j(\bm{y}_{j,n}) 
+ \bm{x}_{i,n+1}\right)
}
\label{eq:ultimodel}
\eeq
In the above formulation, we use an index $n=0,1,\ldots$ to denote the $n$-th time epoch, and an index $i=1,2,\ldots N$ to denote the $i$-th agent of the network. The relevant quantities in~(\ref{eq:ultimodel}) are as follows.

\noindent
{\em ---} The random variable $\bm{y}_{i,n}$ denotes the state of agent $i$ at time $n$. 

\noindent
{\em ---} The random variable $\bm{x}_{i,n}$ represents an input source of randomness or noise, affecting agent $i$ at time $n$.

\noindent
{\em ---} A {\em local nonlinear} coupling between agents $i$ and $j$ at time $n$ is determined by the product $g_i(\bm{y}_{i,n}) h_j(\bm{y}_{j,n})$.

\noindent
{\em ---} 
We will be dealing with {\em weighted directed graphs}, where: $i)$ the graph accounts for the topology linking pairs of nodes; $ii)$ the interaction between pairs of nodes can be {\em directional}, e.g., a coupling effect can exist {\em from} $i$ to $j$, but not {\em from} $j$ {\em to} $i$; and $iii)$ the graph weights are encoded in a {\em combination or interaction matrix} $A$, whose $(i,j)$-th entry, $a_{ij}$, quantifies the strength of interaction in the directional coupling from $i$ to $j$.
We observe from~\eqref{eq:ultimodel} that, if $a_{i j}=0$, then information about the state of agent $j$ does not {\em flow} to agent $i$. Otherwise, if $a_{i j}\neq 0$, then the state of agent $j$ directly impacts the state of agent $i$, provided that the coupling functions $g$ and $h$ do not vanish.

\noindent
{\em ---} $\sigma_i: \mathbb{R}\rightarrow \mathbb{R}$ is a nonlinear {\em invertible} function that plays the role of an {\em activation} function (such as a sigmoidoscopy function). Depending on its particular shape, the function $\sigma_i$ can emphasize or de-emphasize its input values. 
At two extremes, a constant $\sigma_i$ kills the dynamics, whereas a linear $\sigma_i$ is basically transparent to its input. 

Formulations of the type shown in~(\ref{eq:ultimodel}) are often used to model interactions over nonlinear oscillators~\cite{Wang9300,AlderisioetalPRE2017}, or population dynamics and epidemics over networks~\cite{allen1994,KeelingEames2005,topoepidemic2,augustoasilo,augusto_qualitative,KorobeinikovMaini,Korobeinikov2007}. 
For example, in the context of epidemics over networks, the state ${\bm y}_{i,n}$ can model the likelihood of a node $i$ being infected at time $n$ at the individual/microscopical level~\cite{Mieghem}; at the aggregate/macroscopical level, the state can also represent the fraction of infected individuals within community $i$ in an epidemics across communities, as can be established through a thermodynamic or fluid-limit analysis --- see~\cite{Norris,Antunes,AugustoCDC2011,AugustoCDC2012}. 
Likewise, in a general SIR (Susceptible Infected Recovered) formulation, the functions $g_i$ and $h_j$ in~(\ref{eq:ultimodel}) can represent the so-called incidence rate of the infection, whereas the function $\sigma_i$ is simply chosen as the identity function~\cite{KorobeinikovMaini,Korobeinikov2007}. 
Within the aforementioned frameworks, a natural question is whether the underlying network of interactions can be inferred given the evolution of the infection across nodes or communities.

\section{Main Results}
Our main goal is to answer the following {\em identifiability} question: Is it possible to learn the network graph over this class of models? We will see that the answer is in the affirmative. 

To answer, we start by adopting a classic nonlinear regression approach to construct two {\em nonlinear} functions (see~(\ref{eq:zerolag0}) and~(\ref{eq:onelag0}) further ahead for details) that depend on the node measurements. 
More specifically, stacking the observations $\bm{y}_{i,n}$ of all agents at time $n$ into the $N\times 1$ vector $\bm{y}_n$, we will introduce a zero-lag function, $F_0(\bm{y}_n)$, which depends only on the current observation vector $\bm{y}_n$, and a {\em one-lag} function, $F_1(\bm{y}_n,\bm{y}_{n+1})$, which depends on the interaction between observation vectors corresponding to adjacent time epochs.
The expected values of these functions will play an important role:
\beq
\mathcal{F}_0(n)\dfz\E[F_0(\bm{y}_n)],\qquad
\mathcal{F}_1(n)\dfz\E[F_1(\bm{y}_n,\bm{y}_{n+1})].
\eeq
Under suitable conditions on the various nonlinearities $(\sigma, g, h)$, the combination matrix $A$ admits the following closed-form representation:
\beq
\boxed{
A=\mathcal{F}_1(n) \, [\mathcal{F}_0(n)]^{-1}~~~\textnormal{[Our nonlinear case]}
}
\label{eq:mainAinverserel}
\eeq
We shall call this relationship {\em generalized Granger estimator} for the following reason. 
In the special case where $\sigma_i(y)=y$,  $g_i(y)\equiv 1$, and $h_j(y)=y$, the model in~(\ref{eq:ultimodel}) degenerates into the classical {\em linear} model (a.k.a. first-order vector autoregressive model):
\beq
\bm{y}_{i,n+1}=\sum_{j = 1}^N a_{ij} \bm{y}_{j,n} + \bm{x}_{i,n+1}.
\eeq
In this particular case, a well-known representation for $A$ is given by the one-step linear predictor, sometimes called Granger estimator in the context of causal analysis:
\beq
\boxed{
A=R_1(n) [R_0(n)]^{-1}~~~\textnormal{[Classic linear case]}
}
\label{eq:standardGranger}
\eeq
where $R_0(n)=\E[\bm{y}_n \bm{y}_n^{\top}]$ and $R_1(n)=\E[\bm{y}_{n+1} \bm{y}_{n}^{\top}]$ are the zero and one-lag correlation matrices of the samples at time $n$.
The functions $\mathcal{F}_0(n)$ and $\mathcal{F}_1(n)$ depend on the nonlinear functions that characterize the system in~(\ref{eq:ultimodel}), in a way that highlights the role that these nonlinearities play on topology estimation, as we will show in Sec.~\ref{sec:GGE}

The relationship in~(\ref{eq:mainAinverserel}) actually suggests a strategy to estimate the topology. 
However, we see from~(\ref{eq:mainAinverserel}) that $A$ depends upon {\em expected} values, which in turn depend on the knowledge of the distribution of the data. This leads to a circular argument since the distribution would require knowledge of the matrix $A$.
Accordingly, since only a sample path (i.e., one realization of the process) is observed, in order to show that~(\ref{eq:mainAinverserel}) can be useful, we will prove that the aforementioned expected values can be {\em consistently learned from the samples}.
To this aim, it will be critical to establish that the considered dynamical system possesses the following ergodic property (the symbol $\stackrel{\textnormal{a.s.}}{\longrightarrow}$ denotes almost-sure convergence as $n\rightarrow\infty$):
\beqa
\frac{1}{n}\,\sum_{k=0}^{n-1} F_0(\bm{y}_k)\stackrel{\textnormal{a.s.}}{\longrightarrow} \mathcal{F}_0,\quad
\frac{1}{n}\,\sum_{k=0}^{n-1} F_1(\bm{y}_k,\bm{y}_{k+1})\stackrel{\textnormal{a.s.}}{\longrightarrow} \mathcal{F}_1
\label{eq:empiricalF0F1intro}
\eeqa
with:
\beq
\boxed{
A=\mathcal{F}_1 \, \mathcal{F}_0^{-1}
}
\eeq
The main contributions of this work are as follows.
We first answer an {\em identifiability} question, namely, we establish whether graph learning is possible over the considered class of logistic nonlinear dynamical systems.  
To this aim, we exploit a closed-form relationship existing between the interaction matrix $A$ and a pair of functionals of the samples that arise from a solution of a nonlinear regression problem. 
Through this closed-form representation, we characterize the various nonlinearities and attributes that determine the dynamics in~(\ref{eq:ultimodel}), in order to ascertain under which conditions the graph can be learned consistently. 
This characterization is obtained through a set of transparent sufficient conditions, which allows relating the graph identifiability to the physical evolution of the dynamics. 
We note that the method and theory are consistent for arbitrary topologies (whether directed or undirected, and dense or sparse).

Second, to show that the graph is consistently learned as the number of samples grows, we will prove that the consistency result in~(\ref{eq:empiricalF0F1intro}) holds true, through a limiting characterization that builds on powerful results available for Markov chains in the general state space.

Finally, in order to enlarge the range of application, we explore numerically scenarios where some of the conditions used to carry out the technical analysis are relaxed.  
For example, while the aforementioned theoretical results apply in the regime of full observability (i.e., when all nodes of the network are accessible), in Sec.~\ref{sec:partial} we will show some examples that illustrate how the proposed method can work also in the case of {\em partial observability}, along the lines of what has been developed in~\cite{tomo_icassp,tomo,tomo_dsw,tomo_isit,MattaSantosSayedAsilomar2018,MattaSantosSayedISIT2019} for linear models.

\section{Related Work}
The problem of graph learning arises in several disciplines, giving rise to different terminologies and relevant models, including structural equations, structural dynamical systems, graphical or vector autoregressive models. 
In all these models, the evolution of the observables is determined by some form of local interaction between neighboring nodes, and this structure of interaction is encoded in an underlying  network graph. 
In the following, we provide a list of works that are relevant to the present treatment. 

\subsection{Learning Graphical Models}
Graphical models conform to an important and well-studied class of systems in the framework of topology inference. 
In a graphical model, the state of each agent is represented by the realization of a random variable, and a joint distribution among these variables encodes the dependency among the agents, and, hence, encodes the underlying topology. 
Then, inference about the topology is carried out by assuming that independent samples from the joint distribution can be collected.

There are several works on topology inference for specific classes of graphical models, including, e.g., Ising models~\cite{AnandkumarTanHuangWillskyAOS2012, BreslerEfficiently} and Gaussian graphical models~\cite{AnandkumarTanHuangWillskyJMLR2012}.
These works deal with topology inference under the assumption that measurement from all network nodes can be gathered. 
Another relevant framework, especially over large networks, is that of {\em partial observations}. This paradigm corresponds to a challenging inverse problem, where one tries to figure out the graph connecting the observed nodes, despite the latent influence of the unobserved ones. 
With reference to the partial observations setting for graphical models, in~\cite{AnandkumarValluvanAOS2013} a technique to learn the topology over {\em large-girth} graphs (e.g., the bipartite Ramanujan graphs or the random Cayley graphs) is proposed.
In~\cite{ChandrasekaranParriloWillskyAOS2012}, a consistent graph-learning algorithm is proposed under the assumption that the adjacency graph matrix is sparse, and that the error matrix associated to the latent-variables is low-rank. 
In~\cite{BreslerBoltzmann}, an approach based on influence maximization is adopted to establish when the graph learning problem is feasible over the class of restricted Boltzmann machines. 

However, the independence among samples is one fundamental assumption in the framework of graphical models, i.e., it is often assumed that the previous system state does not affect the next state.
For this reason, graphical models do not  natively match the {\em dynamical} system framework that we need here. Likewise, the shape of proper graph estimators can differ substantially from those suited to a dynamical system. 
For example, while in a Gaussian graphical model, the inverse of the correlation matrix (a.k.a. precision matrix) contains full information on the graph topology, the Granger estimator $R_1 R_0^{-1}$ that is the optimal solution for a first-order diffusion model must rely also on the dependency between subsequent samples as encoded in the one-lag correlation matrix $R_1$.

\subsection{Graph Learning with Linear Dynamics}
Many works on graph learning or causal relationship identification focus on linear models, such as diffusion or Vector AutoRegressive (VAR) models~\cite{mateos}. 
These linear systems are of practical interest since they arise in several applications, for example, they are used in economics~\cite{Moneta}; they may represent well the linearized dynamics of more general nonlinear systems~\cite{PhysRevE.lai}; they can describe the dynamics of biological systems~\cite{Sayed} and the operation of distributed algorithms such consensus or diffusion~\cite{MattaSayedCoopGraphSP2018,MattaBracaMaranoSayedTIT2016,MattaBracaMaranoSayedTSIPN2016,KarMouraJSTSP2011}.

The problem of topology inference over linear systems has been actively studied in the past several years. 
A great emphasis lies on exploiting the natural regression formulation that these linear systems exhibit and on reinforcing priors on the network structure (e.g., sparsity, smoothness) when available. 
For example, recent techniques include: spectral-domain techniques based on optimization with sparsity constraints~\cite{pasdeloup}; graph signal processing techniques applied to causal graph processes~\cite{MeiMoura}; directed information graphs~\cite{Kiyavash1} to infer causal dynamics; and approaches based on Wiener filtering to infer the topology~\cite{MaterassiSalapakaTAC2012}.

There are also works addressing topology inference over linear systems under {\em partial} observations. 
There are results established with reference to particular graphs, such as polytrees~\cite{MaterassiSalapakaCDC2012, KiyavashPolytrees}, as well as results for more general graphs~\cite{Geigeretal15, MaterassiSalapakaCDC2015}.
For the case of large-scale graphs,  asymptotic statistical approaches are exploited, where the conditions on the graph are encoded into average summary indicators, such as the connection probability between any two nodes. 
The works~\cite{tomo_icassp,tomo,tomo_dsw,tomo_isit,MattaSantosSayedAsilomar2018,MattaSantosSayedISIT2019} pursue this approach to show that the graph of the observed component can be faithfully reconstructed as the network dimension scales to infinity, under different regimes of connectivity and/or graph models.

\subsection{Graph Learning with Nonlinear Dynamics}
One common approach to tackle the nonlinear problem is to perform some form of linearization of the system. 
In some cases, by a proper change of coordinates, certain dynamics can be represented linearly.
For example, linearization can be obtained by a variational characterization (under a small-noise assumption)~\cite{Ching2017ReconstructingLI}; by suitable augmentation of observable space dimension~\cite{Koopman_Gon}; by appropriate exploitation of the vector-field Jacobian, in conjunction with a compressed sensing method to mitigate the course of dimensionality and the computational complexity~\cite{Nitzane}. 

However, linear or linearized models, while useful under some favorable (e.g., small-noise or small-deviations) conditions, in other cases offer only a convenient approximation of the real dynamics, failing to capture some important aspects thereof.

One relevant class of genuinely nonlinear models is given by nonlinear vector autoregressive models~\cite{nonlinearShen,surveynonlinear}.
In~\cite{nonlinearShen,surveynonlinear}, the nonlinear interaction is modeled through a sum of nonlinear univariate functions of the node variables, belonging to a certain reproducing kernel Hilbert space. Advanced methods are developed to learn the topology, reinforcing prior information in suitably formulated regression problems in order to boost sample-complexity performance.
These methods are flexible enough to incorporate different regularization constraints (such as, e.g., smoothness or sparsity) as well as to cope with the presence of unknown nonlinearities.
While fairly general, the class of dynamical systems treated in these references does not encompass the class of logistic systems.

Another relevant class of nonlinear systems (more closely related to our approach) focuses on proper modeling of the {\em nonlinear coupling between pairs of node variables}.
These models arise across many domains, especially in fundamental physics applications, with one notable model being the Kuramoto model.
In~\cite{Wang9300}, a general class of continuous-time nonlinear dynamical systems of Kuramoto-type is considered to model oscillators.
The vector field of the dynamical law is approximated in a certain complete orthogonal basis of a Hilbert space (e.g., Legendre/Chebyshev polynomials or Fourier series). The coefficients of this expansion encompass information about the topology. 
Linear regression (i.e., linear in the entries of the coefficients) is used to extract the coefficients and hence the topology.

Recent efforts aim at finding proper measures of influence or causality among the agents. 
The underlying structure of the directed network is then generally obtained under appropriate thresholding of the connectivity-measures. 
For example, functional dependency graphs are introduced in~\cite{Kiyavash2} for a fairly broad class of dynamical systems. 
In~\cite{InfTop}, a ``causal information'' measure is proposed to estimate the underlying network of causal relationships. 
In particular, a Kullback-Leibler based measure is devised for a neural network model.

Along the line of the present work, other works have been devoted to identify the underlying network structure of dynamical systems that model natural phenomena such as oscillatory systems or spread of diseases. 
For example, in~\cite{Giardina059865} the object of inference is the phylogenetic tree that accounts also for evolutionary elements concerning the disease spread under study and a Bayesian method is proposed.
In~\cite{topoepidemic2}, a log-MLE estimator is devised to infer the underlying structure. The dynamical model assumed is of SIR (susceptible-infected-recovered) type, driven by a Poisson process. The stochastic dynamical model yields a particular distribution parametrized by (among other parameters) the network structure. 
In ~\cite{AlderisioetalPRE2017} (for Kuramoto type of models) an estimator based on a {\em heuristic influence function} that maps the relative phase difference between nodes' outputs into an estimate of their pairwise connection is proposed.

\textbf{Notation.} We use boldface letters to denote random variables, and normal font letters for their realizations. Matrices
are denoted by capital letters, and vectors by small letters. 
Sets and events are denoted by calligraphic letters. 
We denote by $\P[\mathcal{A}]$ the probability of event $\mathcal{A}$. 
For a random variable $\bm{y}$, the notation $\E[\bm{y}]$ denotes the expected value of $\bm{y}$.
When we say that $\E[\bm{y}]$ exists and is well-defined, we mean that $\E[|\bm{y}|]<\infty$. 
The symbol $\odot$ denotes the Hadamard product.
The symbol $\mathcal{N}_i$ stands for the set of nodes that point to node $i$ in a directed-graph. 
For a vector $v\in\mathbb{R}^N$, we denote by $\|v\|$ a generic vector norm. When dealing with an $N\times N$ matrix $M$, the symbol $\|M\|$ will denote the matrix norm induced by the particular vector norm $\|\cdot\|$, which is defined as:
\beq
\|M\|=\sup_{v\neq 0}\frac{\|M v\|}{\|v\|}.
\eeq
We will be dealing with random variables defined at network nodes and evolving over time. 
The notation $\bm{y}_{i,n}$ will generally denote a (random) variable defined at time $n=0,1,\ldots$, and at node $i\in\{1,2,\ldots,N\}$. With the notation $\bm{y}_n$ we shall denote the vector that collects the variables of all nodes at time $n$, namely,
\beq
\bm{y}_n=[\bm{y}_{1,n},\bm{y}_{2,n},\ldots,\bm{y}_{N,n}]^{\top}.
\eeq

\section{Nonlinear model}

In a networked dynamical system, the state of each agent evolves over time as a result of their peer-to-peer interactions. 
In particular, in the context of discrete-time systems, the state of an agent $i$ at time $n+1$ depends upon its own current state at time $n$ and also on the state of its immediate neighbors at time $n$. The underlying network defining the neighborhood plays critical role in the long-term properties of such dynamical systems. In its most general form, a discrete-time first-order {\em time-homogeneous} {\em continuous state-space} nonlinear stochastic system can be described by the following law (refer to Proposition~$7.6$ in~\cite{kallenberg2002})
\beq
\bm{y}_{i,n+1}=T_i\left({\bm y}_n,\bm{x}_{n+1}\right),
\label{eq:Nmodel}
\eeq
or in vector form
\beq
\bm{y}_{n+1}=T\left({\bm y}_n,\bm{x}_{n+1}\right),
\label{eq:Nmodel}
\eeq
where $T_i\,:\,\mathbb{R}^N\rightarrow \mathbb{R}$, is the vector-field (at node $i$) representing how the local interactions affect the evolution of node $i$; $\{{\bm x}_n\}$ is an i.i.d random process modeling exogenous input/perturbations across nodes over time~$n$. 
The process is time-homogeneous because the law $T$ depends only on the values of the states and of the noise, and does not change over time. The process is with continuous state space because the range of output values can in general vary in $\mathbb{R}^N$.

One characterizing property of such a networked dynamical system is its {\em locality}: the state ${\bm y}_{i,n+1}$ of agent $i$ at time $n+1$, only depends on its own state ${\bm y}_{i,n}$ and the states of the agents within its neighborhood $\{{\bm y}_{j,n}\}_{j\in \mathcal{N}_i}$ at time $n$. In other words, ${\bm y}_{i,n+1}$ is independent of the state of all nodes outside its neighborhood, given the state of the nodes in its neighborhood.
This property is referred to as a {\em local Markov property} and it is naturally induced by the characterizing vector field, $T_i$, which must be sensitive only to the entries associated with the node $i$ and its neighbors in $\mathcal{N}_i$, formally, for any vectors $y\in\mathbb{R}^N$ and $x\in\mathbb{R}^N$, the function $T_i(y,x)$ depends only on $y_i, \{y_j\}_{j\in\mathcal{N}_i}$ and $x$.
The underlying network topology critically determines the evolution of the vector field via this local Markov property. 

In this paper, we are interested in the {\em inverse} problem of inferring the underlying network characterizing the local Markov property of $T$ given the samples $\{{\bm y}_n\}$. It is however hard to devise a universal scheme to extract consistently information about the underlying network of interactions over such a general class of dynamical systems given by~\eqref{eq:Nmodel}. It is obvious that not all models can allow consistent graph learning. 

For this reason, in this work we focus on a subclass of discrete-time continuous-state systems with the vector field defined by~(\ref{eq:ultimodel}), which can be compactly represented in vector form as\footnote{In order to avoid confusion, we note explicitly that the notation $g(\bm{y}_n) \odot A h(\bm{y}_n)$ used in~(\ref{eq:ultimodelvec}) denotes the Hadamard product between $g(\bm{y}_n)$ and the {\em vector} $A h(\bm{y}_n)$.}:
\begin{equation}
\boxed{
\bm{y}_{n+1} = \sigma\left( g(\bm{y}_n) \odot A h(\bm{y}_n) +\bm{x}_{n+1}\right)
}
\label{eq:ultimodelvec}
\end{equation}
having defined the vector-valued functions of vector argument $y\in\mathbb{R}^N$:
\beqa
\sigma(y)&\dfz&[\sigma_{1}(y_1), \sigma_{2}(y_2),\ldots,\sigma_{N}(y_N)]^{\top},\\
g(y)&\dfz&[g_1(y_1), g_2(y_2),\ldots,g_N(y_N)]^{\top},
\\
h(y)&\dfz&[h_1(y_1), h_2(y_2),\ldots,h_N(y_N)]^{\top}.
\eeqa
Throughout this work, we assume that $\{\bm{x}_n\}$ are independent and identically distributed (i.i.d.) random vectors, with zero-mean and finite second moments, and independent of the initial condition $\bm{y}_0$.

Finally, it is worth noting that, since $\sigma$ is an invertible mapping, by setting $\bm{z}_n=\sigma^{-1}(\bm{y}_n)$, the system in~(\ref{eq:ultimodelvec}) can be also represented as:
\beq
\bm{z}_{n+1}=\widetilde{g}(\bm{z}_{n}) \odot A \widetilde{h}(\bm{z}_{n}) + \bm{x}_{n+1},
\label{eq:additivenoisemodel}
\eeq
where we have introduced the composition of functions:
\beq
\widetilde{g}(y)=g(\sigma(y)), ~~ \widetilde{h}(y)=h(\sigma(y)).
\label{eq:composedfunc}
\eeq
The type of model in~(\ref{eq:additivenoisemodel}) is commonly referred to as {\em additive noise model}, and is extensively studied, e.g., in the literature of stochastic dynamical systems~\cite{ChaosFractalsNoise,MCandInvProb,FurtherMarkovBook}. 
Throughout our treatment, we will generally work in terms of the original untransformed model in~(\ref{eq:ultimodelvec}), since working in terms of the composed functions~(\ref{eq:composedfunc}) might obfuscate the role of the different nonlinear functions. 
On the other hand, the representation in~(\ref{eq:additivenoisemodel}) will be useful to prove the technical results in Appendix~\ref{app:ergodicity}, where we will make appeal to existing results pertaining to the additive noise model~\cite{ChaosFractalsNoise,MCandInvProb,FurtherMarkovBook}.

\section{Generalized Granger Estimator}
\label{sec:GGE}
The proper scheme to process the samples with the goal of estimating either the graph-structure or more generally the interaction matrix relies critically on the nature of the samples. Our goal is to establish the proper structure retrieval scheme for the class of networked dynamical systems~\eqref{eq:ultimodel}.

We highlight that for the most general class of networked dynamical systems~\eqref{eq:Nmodel}, one should not expect to have a closed-form expression for the underlying structure in terms of a functional of the samples -- in general, the inference is carried out by indirect means, e.g., as the solution of an optimization problem. By closed-form expression, we mean $A=\mathcal{F}\left({\bm y}_{n+1},{\bm y}_n\right)$, for some functional $\mathcal{F}$ (e.g., expectation) that can be written in closed-form and expressed only in terms of the observable variables.

In what follows, we introduce a {\em weighting function} defined for any $y\in\mathbb{R}$ as:
\beq
\omega(y)\dfz \left[\frac{1}{g_1(y_1)}, \frac{1}{g_2(y_2)},\ldots, \frac{1}{g_N(y_N)}  \right]^{\top}.
\eeq
Since in principle $\omega(y)$ can be singular when $g(y)$ has some nonzero entries, it is useful to introduce the set:
\beq
\mathcal{Z}=\{y\in\mathbb{R}^N: g(y) \textnormal{ has at least one zero entry}\}.
\eeq
Let us also introduce the zero-lag matrix:
\begin{equation}
F_0(\bm{y}_n)\dfz h(\bm{y}_n)h(\bm{y}_n)^{\top},
\label{eq:zerolag0}
\end{equation}
and the one-lag matrix:
\beq
F_1(\bm{y}_{n+1},\bm{y}_n)
\dfz
\left[
\omega(\bm{y}_n)\odot \sigma^{-1}(\bm{y}_{n+1})
\right]
h(\bm{y}_n)^{\top}\,\mathbb{I}(\bm{y}_n\notin\mathcal{Z}),
\label{eq:onelag0}
\eeq
where $\mathbb{I}[\cdot]$ is the indicator function, which takes on the value $1$ if the event under brackets is true, and $0$ otherwise. The indicator has been included in the definition to assign a finite value (zero) to the one-lag matrix when the function $\omega$ is singular.
The next lemma exploits the nonlinear auto-regressive structure of~(\ref{eq:ultimodelvec}) to relate the matrix $A$ to the expected value of two functions of the samples.

\begin{lemma} [Generalized Granger]
\label{theor:GG}
If all the entries in the vector $g(\bm{y_n})$ are nonzero with probability one, and if the following expectations are well-defined:
\beq
\mathcal{F}_0(n)=\E[F_0(\bm{y}_n)],~~
\E\left[
\omega(\bm{y}_n) h(\bm{y}_n)^{\top}
\right]
\label{eq:expectedvalues}
\eeq
then the following expectation is well-defined:
\beq
\mathcal{F}_1(n)\dfz \E[F_1(\bm{y}_{n+1},\bm{y}_n)]
\label{eq:F1welldef}
\eeq
and we have that:
\begin{equation}
\boxed{
\mathcal{F}_1(n)=A\,\mathcal{F}_0(n)
}
\label{eq:ggrangerdirect}
\end{equation}
Moreover, if the matrix $\mathcal{F}_0(n)$ is invertible, we have that:
\begin{equation}
\boxed{
A=\mathcal{F}_1(n)[\mathcal{F}_0(n)]^{-1}
}
\label{eq:ggranger}
\end{equation}
\end{lemma}

\begin{IEEEproof}
From~(\ref{eq:ultimodelvec}) we have the identity:
\beqa
\lefteqn{\omega(\bm{y}_n)\odot \sigma^{-1}(\bm{y}_{n+1})\mathbb{I}[\bm{y}_n\notin\mathcal{Z}]}\nonumber\\
&=&
\left[A h(\bm{y}_n) + \omega(\bm{y}_n)\odot \bm{x}_{n+1}\right]\mathbb{I}[\bm{y}_n\notin\mathcal{Z}].
\label{eq:identity}
\eeqa
Multiplying both sides of~(\ref{eq:identity}) by $h(\bm{y}_n)^{\top}$, from~(\ref{eq:onelag0}) we get:
\beqa
F_1(\bm{y}_{n+1},\bm{y}_n)&=&A h(\bm{y}_n)h(\bm{y}_n)^{\top}\mathbb{I}[\bm{y}_n\notin\mathcal{Z}]\nonumber\\
&+& \bm{x}_{n+1}\odot \omega(\bm{y}_n) h(\bm{y}_n)^{\top}\mathbb{I}[\bm{y}_n\notin\mathcal{Z}].
\label{eq:identity2}
\eeqa
Since by assumption all entries of vector $g({\bm y}_n)$ are nonzero with probability one, we have that $\P[\bm{y}_n\notin\mathcal{Z}]=1$. Thus, by taking expectations of both sides in~(\ref{eq:identity2}), and using the definitions of $F_0(n)$ in~(\ref{eq:expectedvalues}), and of $F_1(n)$ in~(\ref{eq:F1welldef}), we obtain:
\beq
\mathcal{F}_1(n)=A \mathcal{F}_0(n)
+
\E\left[
\bm{x}_{n+1}\odot \omega(\bm{y}_n)h(\bm{y}_n)^{\top}
\right].
\eeq
In view of the definition in~(\ref{eq:zerolag0}), the first term on the RHS is automatically well-defined because we assume integrability of the random variable $F_0(\bm{y}_n)$ --- see the first relationship in~(\ref{eq:expectedvalues}).
As to the second term, from the independence between $\bm{x}_{n+1}$ and $\bm{y}_n$ we can write:
\beqa
\lefteqn{\E\left[
\bm{x}_{n+1}\odot \omega(\bm{y}_n)h(\bm{y}_n)^{\top}
\right]}\nonumber\\
&=&
\E\left[
{\sf diag}(\bm{x}_{n+1})\omega(\bm{y}_n)h(\bm{y}_n)^{\top}
\right]\nonumber\\
&=&
\E\left[
{\sf diag}(\bm{x}_{n+1})\right]
\E\left[
\omega(\bm{y}_n)h(\bm{y}_n)^{\top}
\right]=0,
\eeqa
where the last equality holds since $\mathbb{E}\left[\bm{x}_{n+1}\right]=0$, whereas the second relationship holds in view of the integrability assumption in~(\ref{eq:expectedvalues}), and since $\bm{x}$ has finite first moment~\cite{kallenberg2002}. 
\end{IEEEproof}

\begin{remark}
The solution $\mathcal{F}_1(n)[\mathcal{F}_0(n)]^{-1}$ arises as the classical solution to the following nonlinear regression problem (where $\|\cdot\|_2$ is the Euclidean norm):
\beq
\argmin_{B}\E\left[
\left\|\omega(\bm{y}_n)\odot\sigma^{-1}(\bm{y}_{n+1}) - B h(\bm{y}_n)\right\|_2^2
\right].
\eeq
Likewise, the empirical counterparts of $\mathcal{F}_0(n)$ and $\mathcal{F}_1(n)$:
\beq
\widehat{\mathcal{F}}_0(n)\dfz\frac{1}{n}\,\sum_{k=0}^{n-1} F_0(\bm{y}_k),~~
\widehat{\mathcal{F}}_1(n)\dfz\frac{1}{n}\,\sum_{k=0}^{n-1} F_1(\bm{y}_k,\bm{y}_{k+1})
\label{eq:empirF0F1}
\eeq
would give $\widehat{\mathcal{F}}_1(n)[\widehat{\mathcal{F}}_0(n)]^{-1}$ as the solution of the least-squares fitting problem:
\beq
\argmin_{B}
\sum_{k=0}^{n-1} \|\omega(\bm{y}_k)\odot\sigma^{-1}(\bm{y}_{k+1}) - B h(\bm{y}_k)\|_2^2.
\label{eq:LSfit}
\eeq
~\hfill$\square$
\end{remark}

The relationships in Lemma~\ref{theor:GG} are basically obtained through straightforward algebra. 
However, there are two fundamental issues that have been overlooked so far. 

First, the relationship in~(\ref{eq:ggranger}) contains {\em expected} values. These expected values are obviously unknown (since they depend on the unknown distribution of the samples, which in turn depends on the object of the topology inference, the unknown matrix $A$). Nevertheless, Eq.~(\ref{eq:ggranger}) would become very useful if we can show that the matrix functions $\mathcal{F}_0(n)$ and $\mathcal{F}_1(n)$ can be {\em consistently estimated from the samples}, e.g., using~(\ref{eq:empirF0F1}). This analysis requires addressing issues of stability and ergodicity, and will be carried out in Sec.~\ref{sec:ergodicity}.~\hfill$\square$

Second, the statement of Lemma~\ref{theor:GG} is in a sense optimistic, since it relies on two generic assumptions like ``{\em assume this is well-defined}'' and ``{\em assume this is invertible}''. 
More precisely, for the direct relationship~(\ref{eq:ggrangerdirect}) to hold, one needs that the expected values in~(\ref{eq:expectedvalues}) are well-defined. 
Once this condition is ascertained, one more condition is needed. Indeed, to get the fundamental {\em inverse} relationship~(\ref{eq:ggranger}), which allows retrieving the object of topology inference, $A$, from the zero-lag and one-lag functions, one needs $\mathcal{F}_0(n)$ to be invertible.~\hfill$\square$

We conclude that the inference problem treated in this work is not tractable for all possible functions $(\sigma, g, h)$ and statistical laws for the noise $\bm{x}$ characterizing the dynamical system.
That is why establishing when the aforementioned conditions are met (we will provide next a set of sufficient conditions for that) is critical to establish whether or not full information about the topology is contained in the samples.
Moreover, it is also relevant to understand the practical meaning of these assumptions in connection to the properties of the networked dynamical system under consideration. The forthcoming sections address all these fundamental concerns.

\subsection{Existence of $\mathcal{F}_0(n)$ and $\mathcal{F}_1(n)$}

Examining~(\ref{eq:ultimodel}), we see that the map $g$ controls the flow of information among nodes. 
To give (an extreme) example, if $g\equiv 0$ then the information about the state of the neighboring agents does not flow or in particular, the network information (that is entailed in $A$) is lost. It is further natural to expect that if $g$ is generally {\em too small} then, even though information is technically flowing, it is hard to extract it. 

Thus, first of all, we have to guarantee that the inverse vector $1/g$ is almost surely well defined, i.e., that the probability that $g$ has some zero entries is zero. One critical difficulty here is that $g$ is a function of $\bm{y}_n$, and that the distribution of $\bm{y}_n$ depends in some intricate way on the evolution of the dynamical system. 

\begin{figure*}[t]
\centering
\includegraphics[width=47mm]{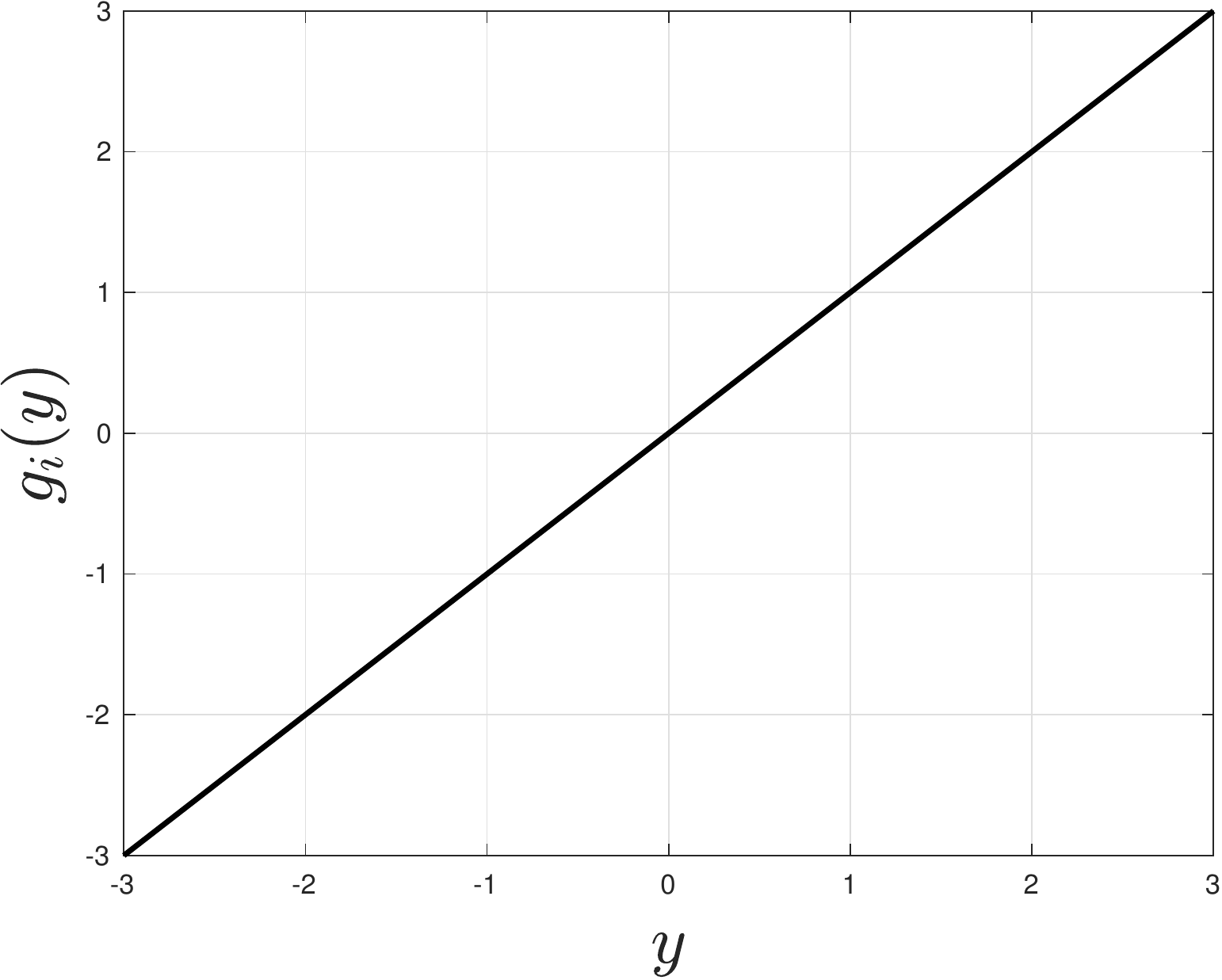}
\includegraphics[width=55mm]{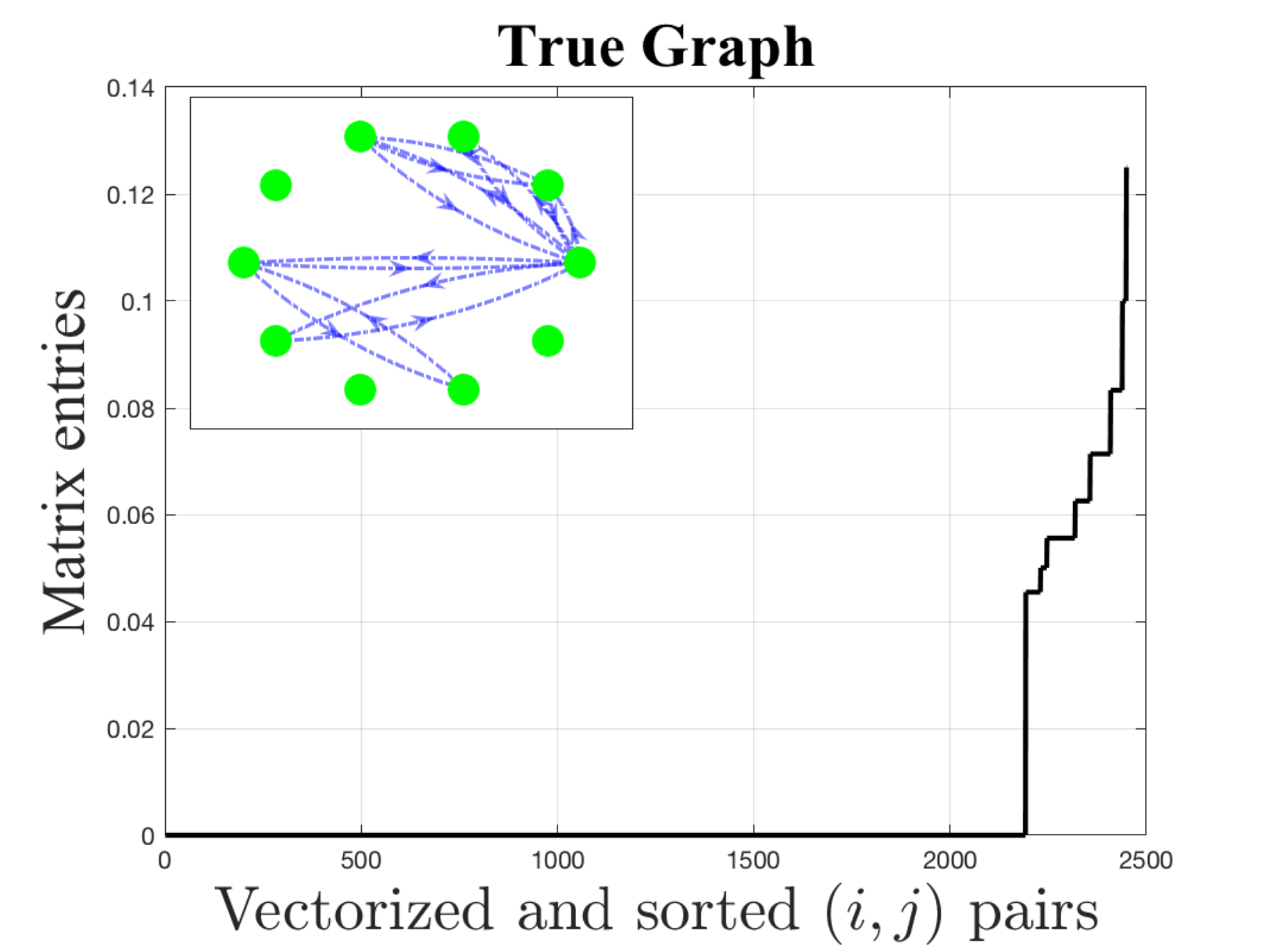}
\includegraphics[width=55mm]{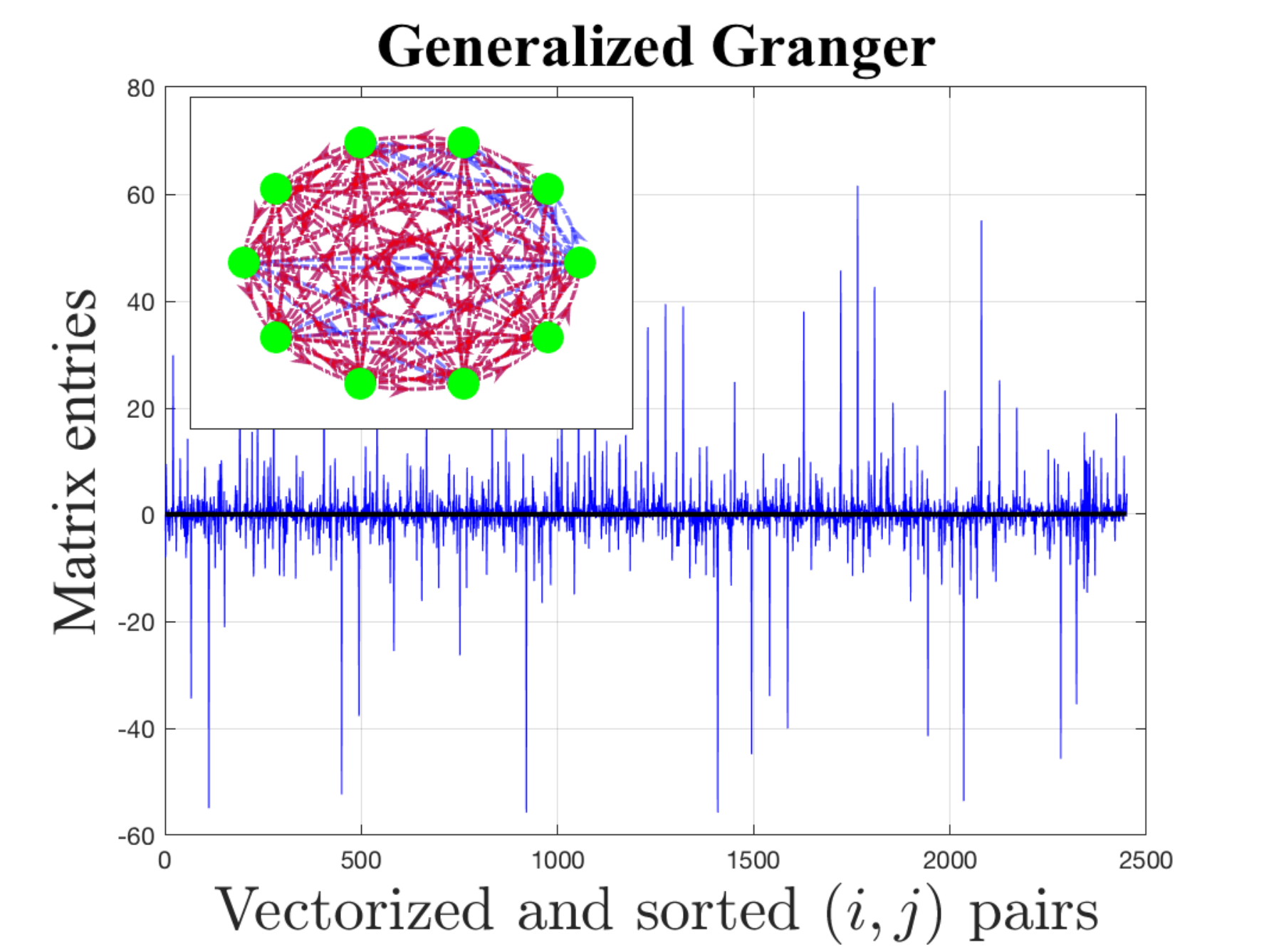}
\caption{An example where the shape of $g$ does impair integrability, resulting in a singular estimator of the graph.}
\label{fig:h1sing}
\end{figure*}

Such physical property reflects into the following mathematical requirement. Since the matrix function $\mathcal{F}_1(n)$ in~(\ref{eq:expectedvalues}) contains the inverse vector function $1/g$, one must guarantee that the latter behaves properly in terms of integrability. 
One example that shows how the result of Lemma~\ref{theor:GG} needs to be carefully checked, is shown in Fig.~\ref{fig:h1sing}, where we used the simple linear function $g_i(y)=y$ for all $i=1,2,\ldots,N$. We evaluate empirically the matrix functions $\mathcal{F}_0(n)$ and $\mathcal{F}_1(n)$, and we observe that they basically blow up.  

The next assumption formally sets a condition on the aforementioned aspect.

\begin{assumption}[Integrability conditions]
\label{assum:oneoverh1}
We assume bounded moments:
\beq
\E[\|\bm{y}_0\|^2]<\infty,~~\mathbb{E}[\|\bm{x}\|^2]<\infty,
\eeq
for the initial condition~$\bm{y}_0$ and the noise term $\bm{x}$.
We assume also that:
\beq
\E[\|\omega(\sigma(\bm{x}+c))\|^2]\leq K,~~\forall c\in\mathbb{R}^N,
\label{eq:oneoverh1boundK}
\eeq
and for some constant $K$.~\hfill$\square$
\end{assumption}

Moreover, since there are several functions involved in the definition of $\mathcal{F}_1(n)$ as well as in the definition of $\mathcal{F}_0(n)$, we need a set of sufficient conditions to guarantee the existence of the latter two matrix functions.

\begin{assumption} [Regularity of the nonlinearities] 
\label{assum:Lipschitz}
We assume that $\sigma$ is a diffeomorphism\footnote{Actually, to prove Proposition~\ref{theor:integrability} we need just $\sigma$ to be continuous and invertible.}.
We assume that $g$ and $h$ are continuous functions.
Furthermore, let us introduce the diagonal matrix:
\beq
D_{g}(y)\dfz{\sf diag}(g(y)).
\eeq
We assume the following conditions on the nonlinearities $(\sigma, g, h)$:
\beqa
\|\sigma(y)\|&\leq& \alpha_{\sigma}\|y\| +\beta_{\sigma},\label{eq:newLipschitzlikesigma}\\
\|D_{g}(y)\|&\leq& \alpha_g\|y\|^p + \beta_g,\label{eq:newLipschitzlikeh1}\\
\|h(y)\|&\leq& \alpha_h\|y\|^q + \beta_h.\label{eq:newLipschitzlikeh0}
\eeqa
for some nonnegative constants $\alpha_{\sigma},\alpha_g,\alpha_h,\beta_{\sigma},\beta_g,\beta_h$
with $p\geq0$, $q\geq 0$, and $p+q=1$.~\hfill$\square$
\end{assumption}
For instance, a condition like~(\ref{eq:newLipschitzlikesigma}) holds when $\|\sigma(y)\|$ grows not faster than $\|y\|$ outside some compact set. Indeed, in this case the constant $\beta_{\sigma}$ can be given by the maximum of $\|\sigma(y)\|$ inside that compact set.

Likewise, when some of the functions $\sigma, h, g$ are bounded, we assume the pertinent $\alpha$-constants equal to zero. 

We have the following result.
\begin{proposition}[Existence of $\mathcal{F}_0(n)$ and $\mathcal{F}_1(n)$]
\label{theor:integrability}
Under Assumptions~\ref{assum:oneoverh1} and~\ref{assum:Lipschitz}, the expectations
\beqa
\mathcal{F}_0(n)&=&\E[h(\bm{y}_n)h(\bm{y}_n)^{\top}],\nonumber\\
\mathcal{F}_1(n)&=&\E\left[
\left(
\omega(\bm{y}_n)\odot \sigma^{-1}(\bm{y}_{n+1})
\right)
h(\bm{y}_n)^{\top}
\right]
\eeqa
are well-defined, and the random variable $g(\bm{y}_n)$ has all nonzero entries with probability $1$.
\end{proposition}
\begin{IEEEproof} The proof is given in Appendix~\ref{app:integrability}.
\end{IEEEproof}

\begin{figure*}[t]
\centering
\[
\begin{array}{cc}
\hspace*{-5mm}
\includegraphics[width=65mm]{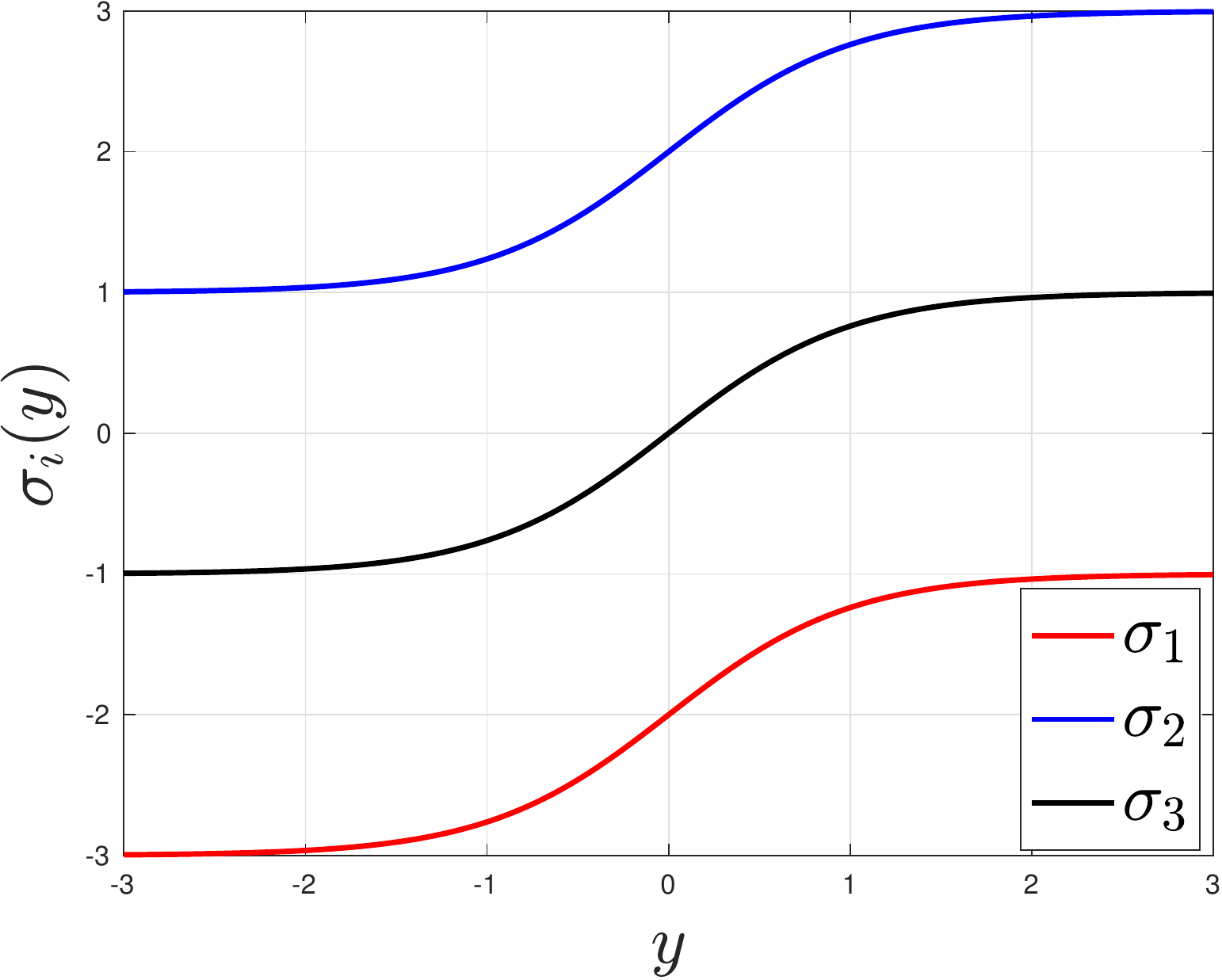}
\hspace*{9mm}
\includegraphics[width=66mm]{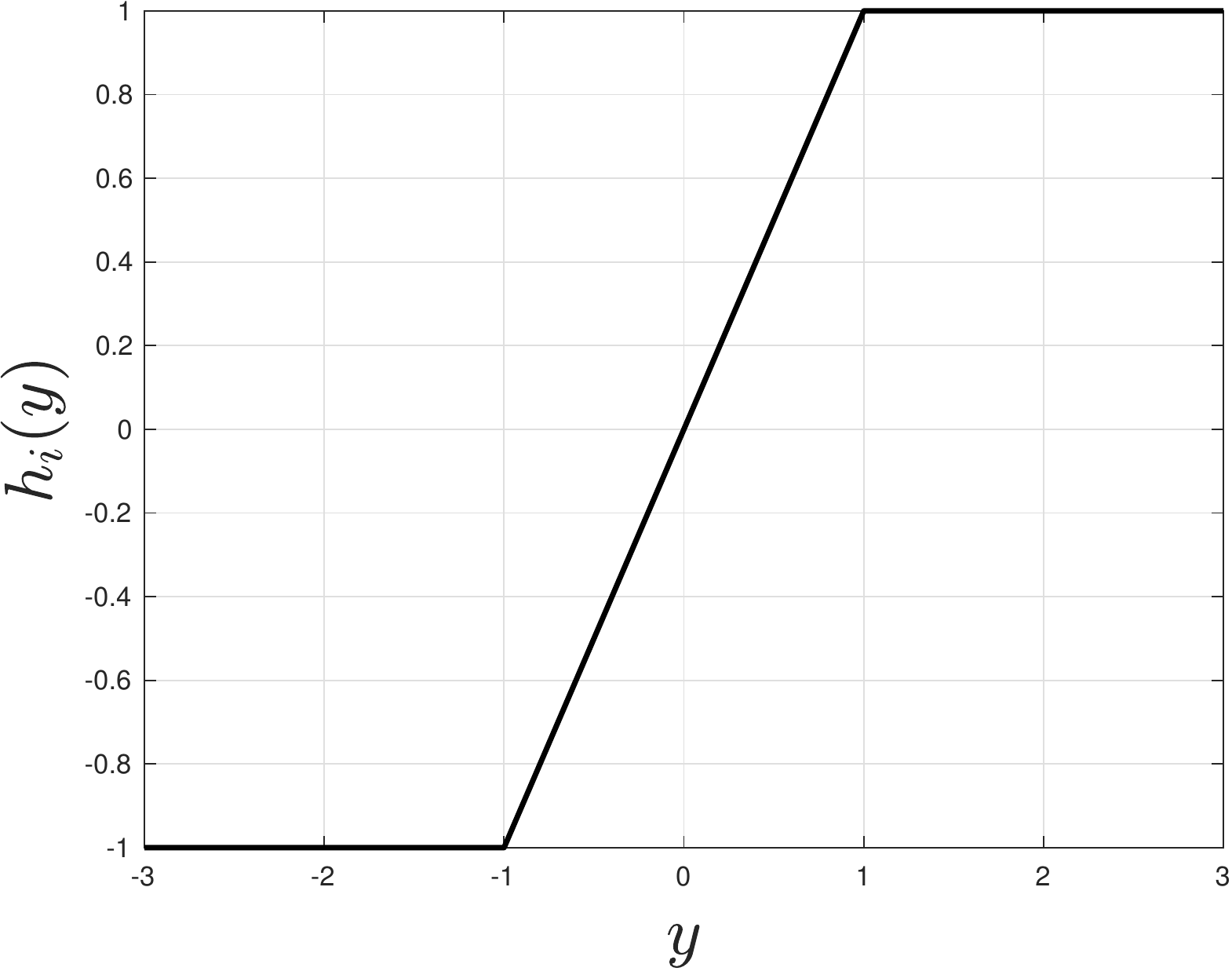}\\
\includegraphics[width=75mm]{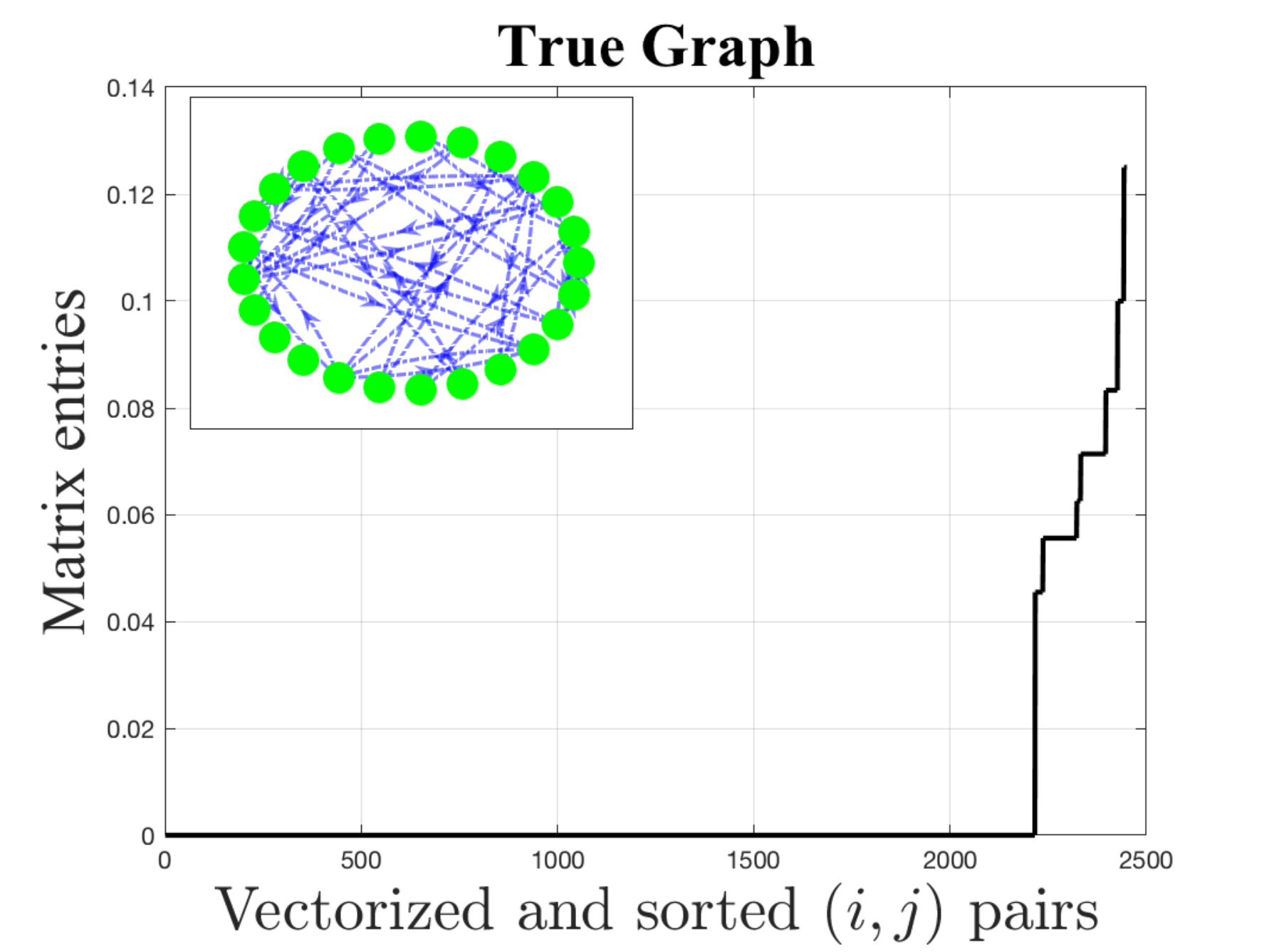}
\includegraphics[width=75mm]{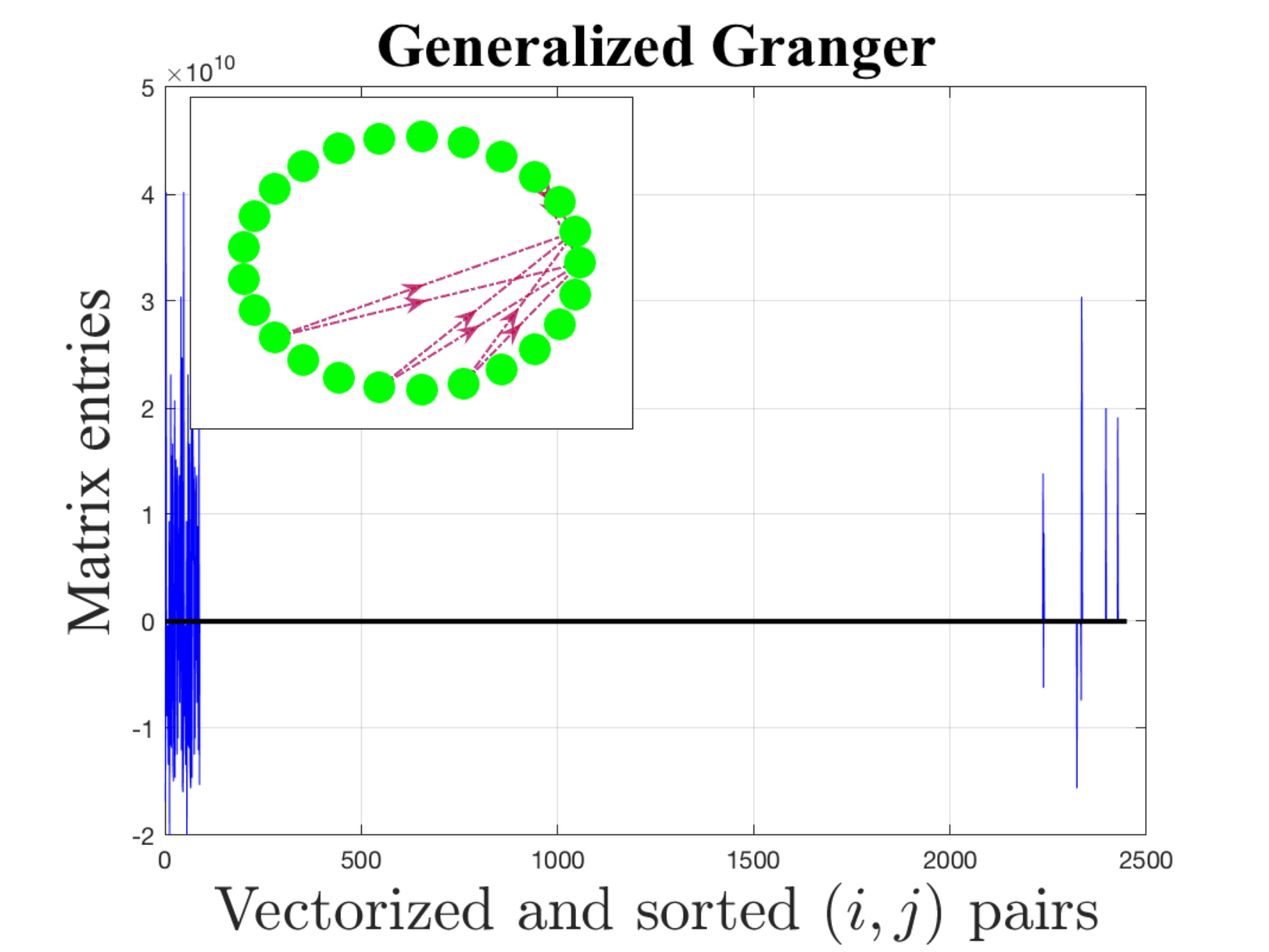}
\end{array}
\]
\caption{An example where the shape of $h$ does impair invertibility of $\mathcal{F}_0(n)$, resulting in a singular estimator of the graph.}
\label{fig:h2sing}
\end{figure*}

\subsection{Invertibility of $\mathcal{F}_0(n)$}
A critical requirement for retrieving $A$ from the matrix functions $\mathcal{F}_0(n)$ and $\mathcal{F}_1(n)$ is invertibility of $\mathcal{F}_0(n)$. However, it is important to remark that there might be dynamical systems where this condition can be violated.

In Fig.~\ref{fig:h2sing}, we show an example where the function $\sigma$ has different behaviors for different agents. In particular, for all agents we have that the component-wise functions have a hyperbolic tangent shape. For agents $1$ and $2$, however, they are shifted (upward and downward, respectively). 
The function $g$ is chosen so as to fulfill the conditions for integrability.
The function $h$ is the limiter (i.e., linear saturating function) displayed in Fig.~\ref{fig:h2sing}. 
We can give to these nonlinearities some physical meaning. The nonlinear shape is typical of several applications (for instance, neural networks). The saturation effects present both in $\sigma$ and $h$ are as well typical. 
For instance, the saturation in $h$ takes on the practical meaning of limiting the interaction when the sample $\bm{y}_{i,n}$ is too large. This behavior can be present, e.g., in a social learning problem where the agents might use saturation to filter observations that look like outliers.

In the example reported in Fig.~\ref{fig:h2sing}, we have computed empirically the matrices $\mathcal{F}_0(n)$ and $\mathcal{F}_1(n)$. In particular, we have verified that $\mathcal{F}_0(n)$ is not invertible, which gives rise to the singular behavior of the topology estimator observed in the lowermost-rightmost panel.
This behavior is due to the fact that the vector function $h$ (uppermost-rightmost panel) becomes constant along the two coordinates corresponding to nodes $1$ and $2$, since the output range of $\sigma_1$ and $\sigma_2$ (uppermost-leftmost panel) forces $h_1$ and $h_2$ to operate only in the saturation region (i.e., $y<-1$ for node $1$, and $y>1$ for node $2$). This yields singularity of the matrix:
\beq
\mathcal{F}_0(n)=\E[h(\bm{y}_n) h(\bm{y}_n)^{\top}].
\eeq
Indeed, since the matrix $\E[h(\bm{y}_n) h(\bm{y}_n)^{\top}]$ is always positive {\em semi}-definite, in order to grant invertibility we must exclude the condition that there exists some (deterministic) vector $v$ such that:
\beq
v^{\top} \E[h(\bm{y}_n) h(\bm{y}_n)^{\top}] v=0.
\eeq
However, the above condition would correspond to say that:
\beq
h(\bm{y}_n)^{\top} v=0~~\textnormal{almost surely},
\label{eq:vorthogas}
\eeq
which basically means that $h(\bm{y}_n)$ must not be a low-rank map. 
The following assumption (which is, e.g., usually encountered in ergodic theory~\cite{ChaosFractalsNoise}) makes this statement precise.

\begin{assumption} [Non-singularity of response $h$]
\label{assum:assumh2}
Let $\mu_{\textnormal{Leb}}$ be the Lebesgue measure in $\mathbb{R}^N$. 
We assume that $h\,:\,\mathbb{R}^N\rightarrow\mathbb{R}^N$ is such that
\begin{equation}\label{eq:nondegenerate}
\mu_{\textnormal{Leb}}(h^{-1}(\mathcal{A}))=0 \textnormal{ for all sets } \mathcal{A} \textnormal{ such that } \mu_{\textnormal{Leb}}(\mathcal{A})= 0.
\end{equation}
In words, this assumption states that if the input set $\mathcal{A}$ {\em has full-dimension}, then its image $h(\mathcal{A})$ is non-degenerate as well. 
It can be verified that transformations that fulfill property~\eqref{eq:nondegenerate} are: linear full-rank maps, open maps, diffeomorphisms, and differentiable maps with non-singular Jacobian almost everywhere~\cite{ChaosFractalsNoise}.

On the other hand, a constant map does not have this property or, more generally, a low-rank linear map (e.g., a projection onto a lower-dimensional subspace) does not have this property as the image of any set necessarily lies in a lower-dimensional subspace (orthogonal to the kernel of the linear application).~\hfill$\square$
\end{assumption}

\begin{proposition}[Invertibility of $\mathcal{F}_0(n)$]
\label{theor:invertibility}
Under Assumptions~\ref{assum:oneoverh1}--~\ref{assum:assumh2}, and if $\sigma$ is a diffeomorphism, then the matrix 
\beq
\boxed{
\mathcal{F}_0(n)=\E[h(\bm{y}_n) h(\bm{y}_n)^{\top}]
}
\eeq
is invertible for all $n\in\mathbb{N}$.
\end{proposition}
\begin{IEEEproof} The proof is given in Appendix~\ref{app:invertibility}.
\end{IEEEproof}

\section{Consistent Graph Learning}
\label{sec:ergodicity}
Lemma~\ref{theor:GG} states that the interaction matrix $A$ can be obtained in terms of two matrix functions, namely, the functions $\mathcal{F}_0(n)$ and $\mathcal{F}_1(n)$. 
However, this property alone would not be sufficient to guarantee that we can reliably estimate the topology from the samples $\{\bm{y}_n\}$. For this to be possible, we should demonstrate that the aforementioned functions {\em can be consistently learned from the samples}. The requirement of consistency means that we should be able to converge to the desired matrix functions as the number of available samples grows.  
This property would yield a practical scheme to estimate the interaction matrix $A$ (and hence its support) from the samples. 

In other words, we are requiring the system to be stable and ergodic. 
And indeed, inference problems over dynamical systems often rely on the stability of the system: it is hard to perform faithful inference over systems that blow up. 

Technically speaking, the model in~(\ref{eq:ultimodelvec}) corresponds to a Markov chain with a {\em general state space}, since, for instance, when the noise component $\bm{x}$ is absolutely continuous, the chain can walk over a {\em continuous} space~\cite{MeynTweedie}.
For such type of Markov chains, the limiting results (e.g., stability and/or ergodicity) are much more involved than those corresponding to the classic case of finite or discrete state space. In the following, we will appeal to powerful results to prove that our dynamical system is in fact ergodic, which would be a critical property to enable consistent topology learning.

We have the following result.
\begin{proposition}[Consistent Graph Learning]
\label{theor:ergodicity}
Assume that Assumptions~\ref{assum:oneoverh1}--\ref{assum:assumh2} are fulfilled.
Assume further that the noise $\bm{x}_n\in\mathbb{R}^N$ is absolutely continuous with almost-everywhere (with respect to the Lebesgue measure in $\mathbb{R}^N$) positive density. 
Let us introduce the stability constant:
\beq
\kappa_s\dfz\left\{
\begin{array}{ll}
\alpha_{\sigma}\alpha_g\alpha_h\|A\|,~~~~~~~\textnormal{if $p>0$ and $q>0$},\\
\alpha_{\sigma}\alpha_g\beta_h\|A\|,~~~~~~~\textnormal{if $p=1$ and $q=0$},\\
\alpha_{\sigma}\alpha_h\beta_g\|A\|,~~~~~~~\textnormal{if $p=0$ and $q=1$}.
\end{array}
\right.
\label{eq:kappadef}
\eeq
Then, if $\kappa_s<1$, we have that:
\beq
\boxed{
A=\mathcal{F}_1 \, \mathcal{F}_0^{-1}
}
\eeq
where
\beq
\mathcal{F}_0=\lim_{n\rightarrow\infty} \mathcal{F}_0(n),~~
\mathcal{F}_1=\lim_{n\rightarrow\infty} \mathcal{F}_1(n),
\label{eq:F0F1expeconv}
\eeq
and:
\beqa
\widehat{\mathcal{F}}_0(n)&=&\frac{1}{n}\,\sum_{k=0}^{n-1} F_0(\bm{y}_k)\stackrel{\textnormal{a.s.}}{\longrightarrow} \mathcal{F}_0,\label{eq:asconvtheorergF0}\\
\widehat{\mathcal{F}}_1(n)&=&\frac{1}{n}\,\sum_{k=0}^{n-1} F_1(\bm{y}_k,\bm{y}_{k+1})\stackrel{\textnormal{a.s.}}{\longrightarrow} \mathcal{F}_1.
\label{eq:asconvtheorergF1}
\eeqa
\end{proposition}
\begin{IEEEproof} The proof is given in Appendix~\ref{app:ergodicity}.
\end{IEEEproof}

Proposition~\ref{theor:ergodicity} provides {\em consistency} (i.e., faithful estimation as the number of samples grows) of the generalized Granger within the scope of the class of nonlinear dynamical systems~\eqref{eq:ultimodel}. 
Based on this result, we are now in the position of proposing the following graph learning algorithm to reconstruct the underlying (directed) graph from each realization (or sample-path) of the nonlinear system.

\begin{algorithm}\label{tb:algorithm}
    \caption{Empirical Generalized Granger (EGG)}
  \begin{algorithmic}[1]
    \REQUIRE Samples~$\left\{y_k\right\}$, for time epochs $k=0,1,\ldots,n$.
    \\ 
    \hspace*{16pt}Dynamical-law triple $(\sigma, g, h)$
    \vspace*{5pt}
    \ENSURE $\widehat{A}(n)=$ Estimate of $A$
    \STATE $\widehat{\mathcal{F}}_0(n) = \displaystyle{ \frac{1}{n}\sum_{k=0}^{n-1} h(y_k)h(y_k)^{\top}}$
    \STATE $\widehat{\mathcal{F}}_1(n) =  \displaystyle{\frac{1}{n}\sum_{k=0}^{n-1} 
    \left[\omega(y_k)\odot \sigma^{-1}(y_{k+1})\right] h(y_k)^{\top}}$
    \STATE $\widehat{A}(n) = \widehat{\mathcal{F}}_1(n)\left[\widehat{\mathcal{F}}_0(n)\right]^{-1}$
    \end{algorithmic}
\end{algorithm}

One special comment is deserved by the last step of the algorithm. 
Once we estimate the combination matrix $A$, we need to reconstruct its support graph. 
However, due to finiteness of the number of samples, also the zero entries in $A$ would result in some (possibly small) nonzero entries. Accordingly, in the last step of the algorithm we apply a clustering algorithm (here the $k$-means with $k=2$) to the entries of the estimated interaction matrix $\widehat{A}$. 
Such clustering algorithm is aimed at devising a boundary between causality and non-causality among agents, i.e., to provide an automated {\em data-driven} classification threshold: $j$ affects $i$ if the entry $\widehat{a}_{ij}$ assumes a high(er) value, otherwise, if $a_{ij}$ assumes a weak(er) value then $j$ is deemed as not affecting $i$.

In order for the clustering algorithm to work properly, one might question that the zero/nonzero entries of $A$ should inherently possess some clustering property. 
Indeed, such form of clustering has been proved to hold (for sufficiently large network sizes) when the underlying graph is an Erd\H{o}s-R\'enyi random graph, for a relevant class of combination matrices~\cite{tomo_icassp,tomo,tomo_dsw,tomo_isit,MattaSantosSayedAsilomar2018,MattaSantosSayedISIT2019}.

\section{Illustrative Examples}

We now present a number of examples aimed at illustrating the validity of the theoretical analysis conducted so far.
In what follows, the {\em ground-truth combination matrix}, $A$, is constructed through the following steps. 
First, we generate a realization of a {\em binomial graph} with connection probability $p$, which means that any arrow (i.e., directed edge) exists with probability $p$, and independently from all the other edges~\cite{BollobasRandom}. 

Once the underlying graph has been generated, the combination weights are assigned to each arrow of the graph to yield the combination matrix $A$. We will consider the following assignment rule (a.k.a. uniform averaging rule):
\begin{equation}\label{eq:unifave}
a_{ij}=\left\{\begin{array}{ll} 
\rho\,\displaystyle{\frac{g_{ij}}{d_i}}, & \mbox{ for }i\neq j\\
\\
\rho - \sum_{k\neq i} a_{ik}, & \mbox{ for }i=j\end{array}\right.,
\end{equation}
where $g_{ij}\in\{0,1\}$ is equal to $1$ only if there is a directed edge from $j$ to $i$, and where~$d_i=|\mathcal{N}_i|$ is the in-degree or number of agents directly influencing agent $i$. 

\begin{figure*}[t]
\centering
\[
\begin{array}{cccc}
\hspace*{-5mm}
\includegraphics[width=38mm]{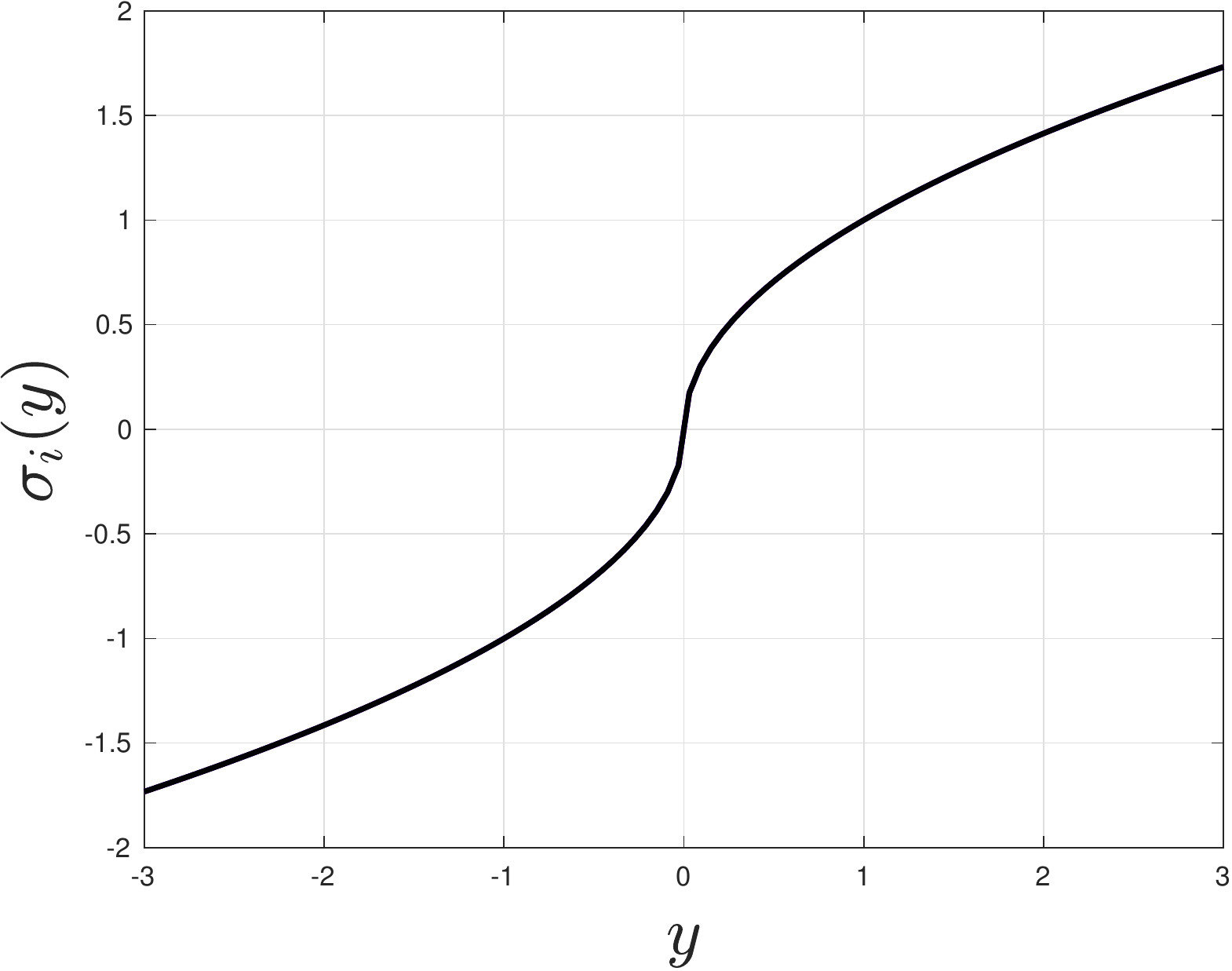}&
\hspace*{-10mm}
\includegraphics[width=38mm]{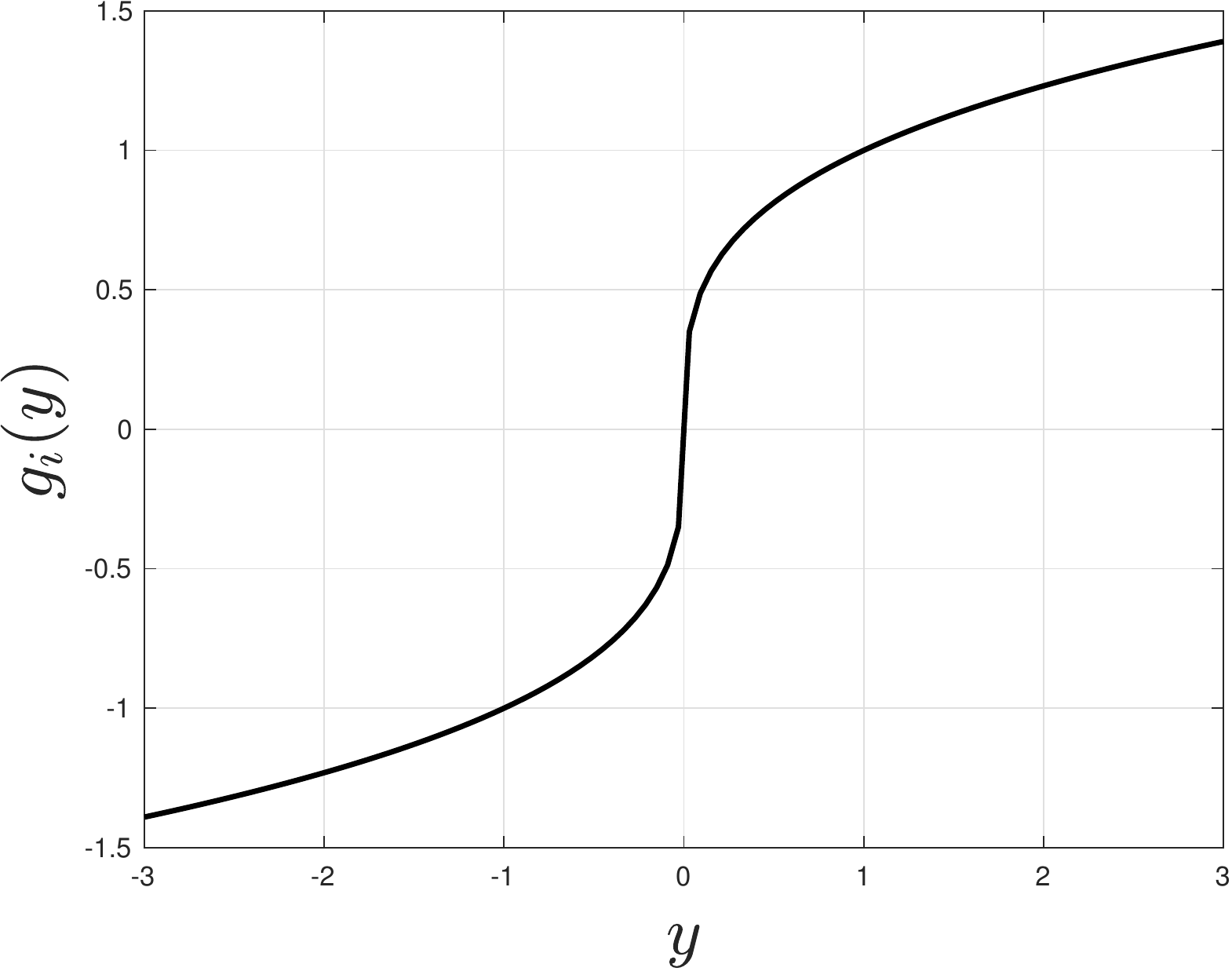}&
\hspace*{-10mm}
\includegraphics[width=38mm]{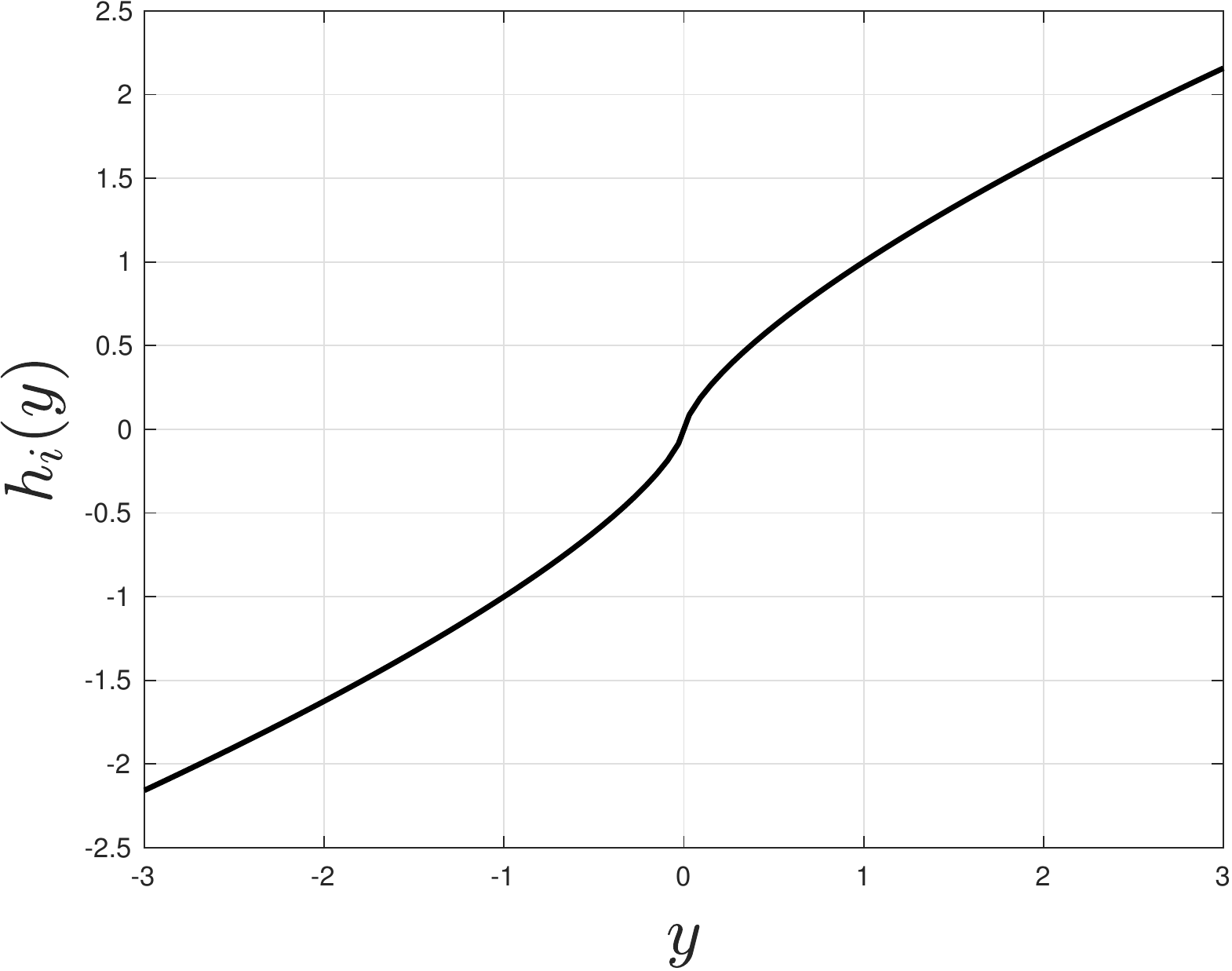}&
\hspace*{-8mm}
\includegraphics[width=44mm]{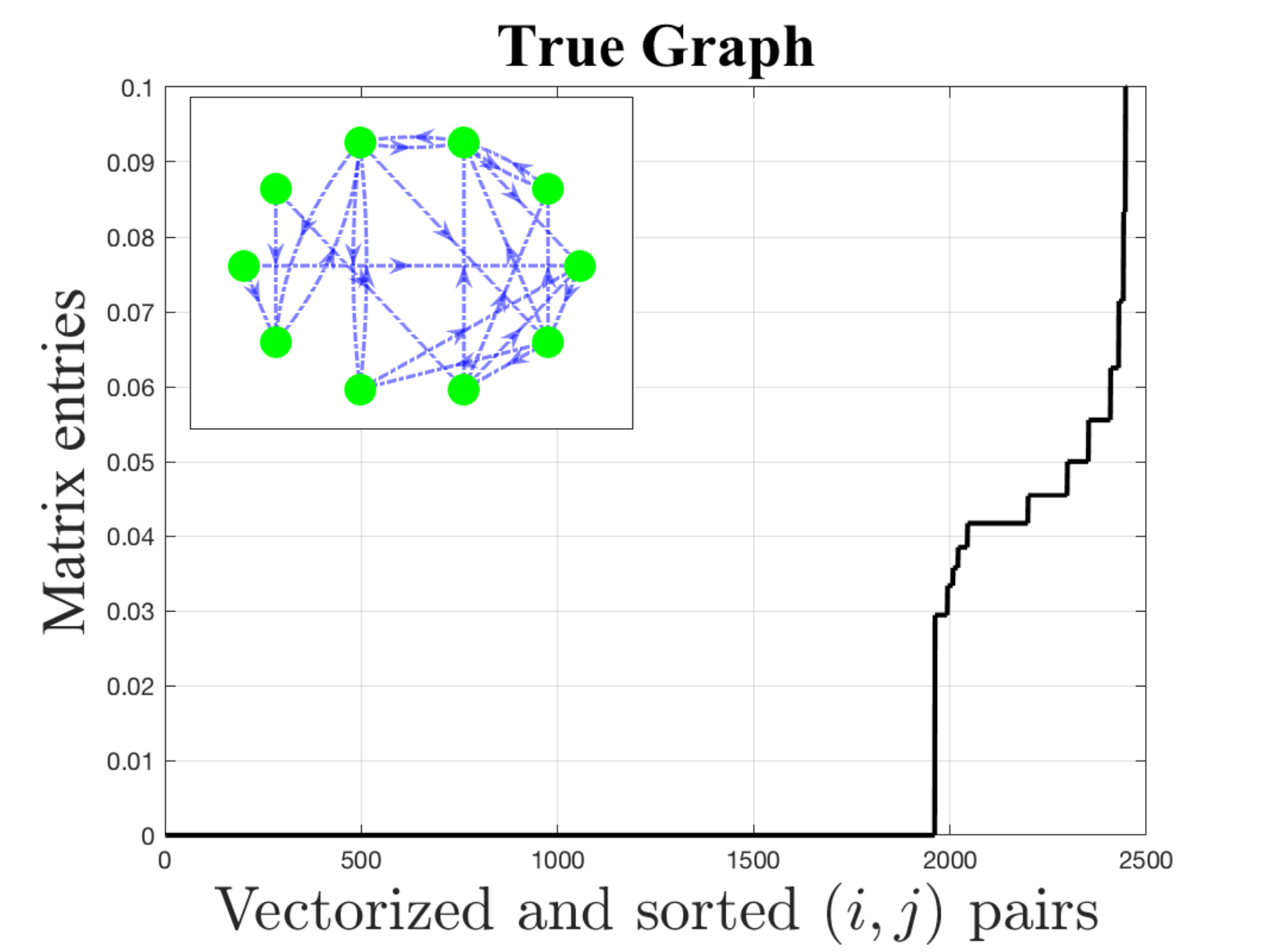}\\
\hspace*{-2mm}
\includegraphics[width=44mm]{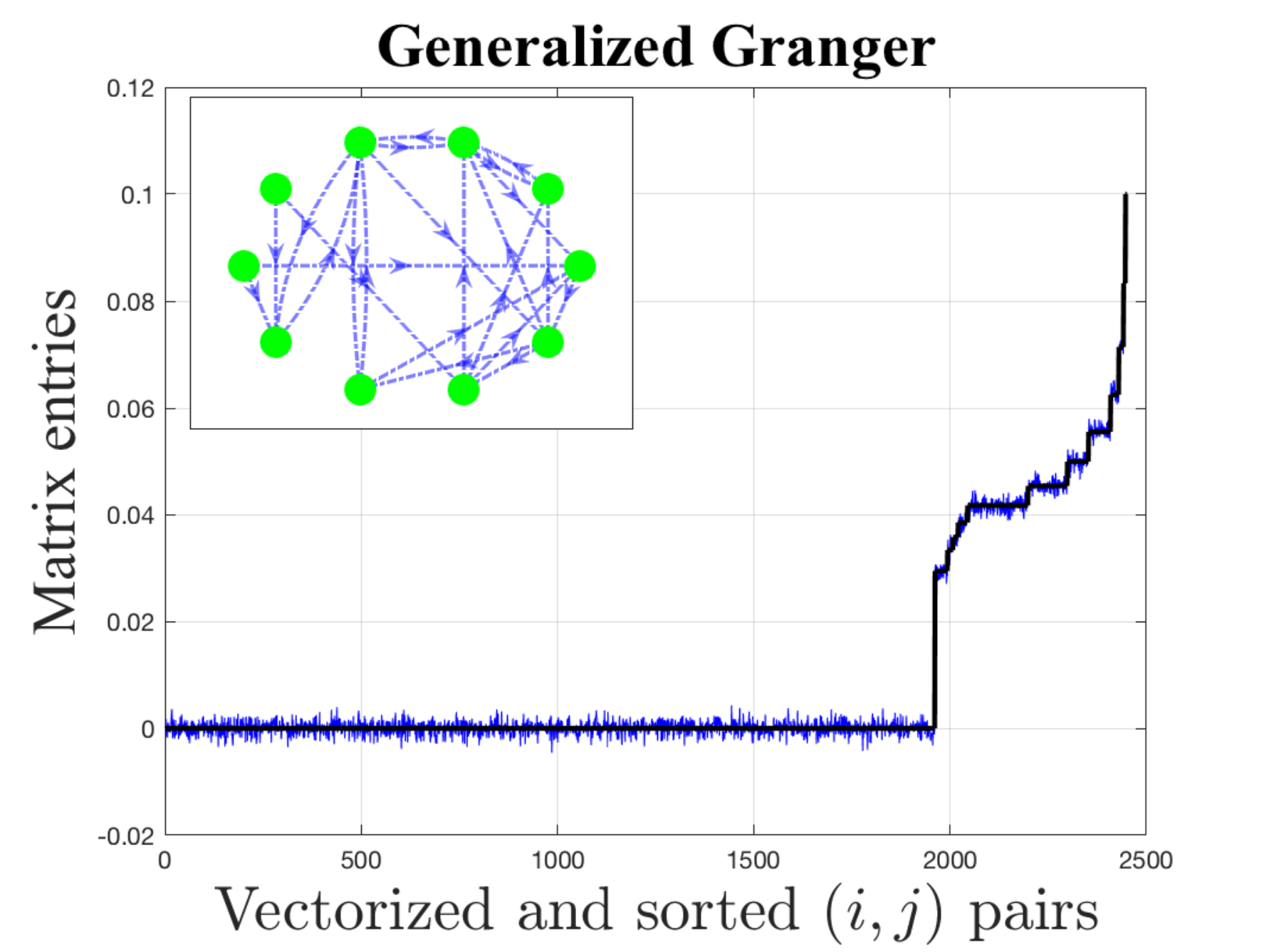}&
\hspace*{-7mm}
\includegraphics[width=44mm]{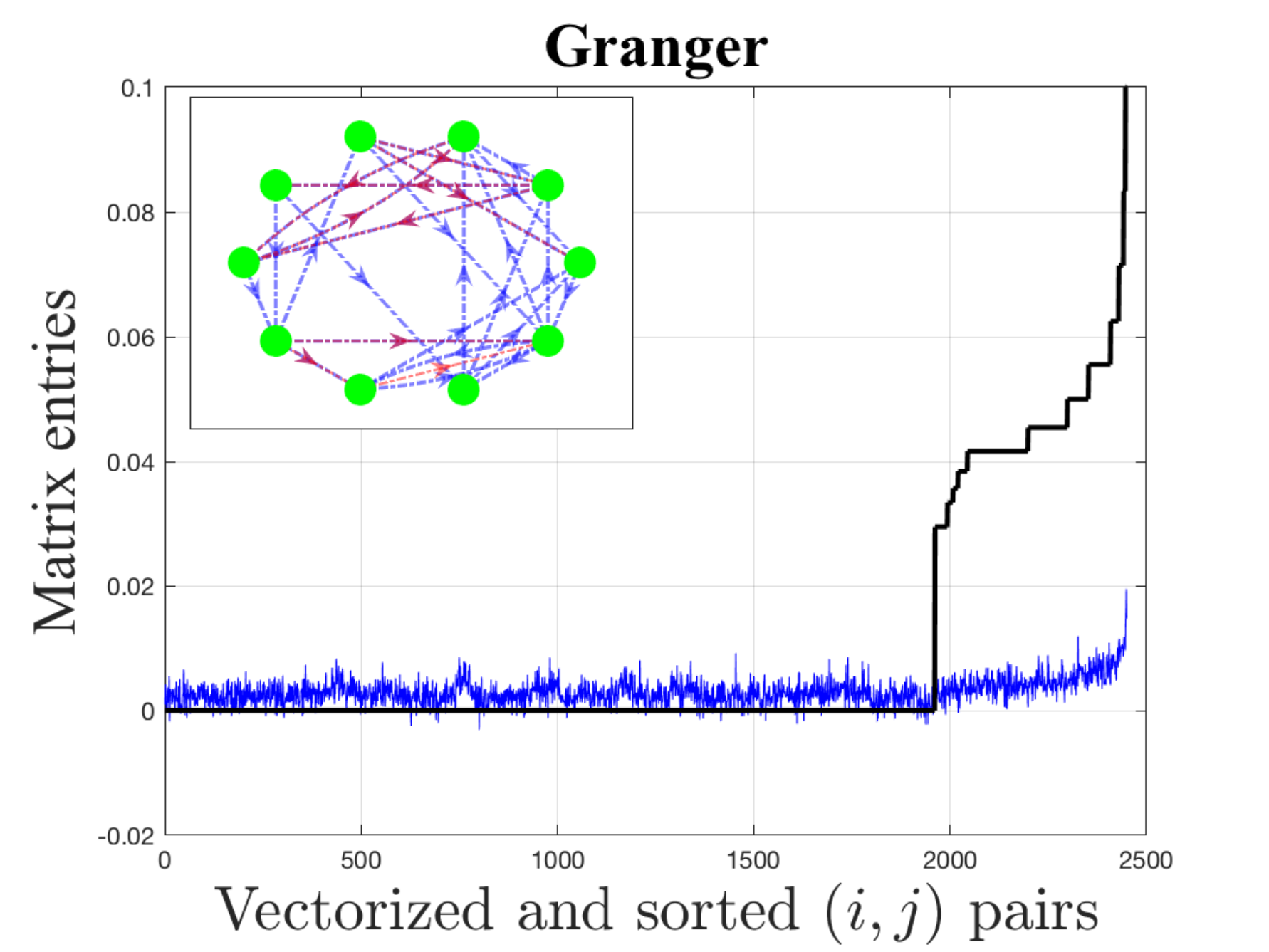}&
\hspace*{-7mm}
\includegraphics[width=44mm]{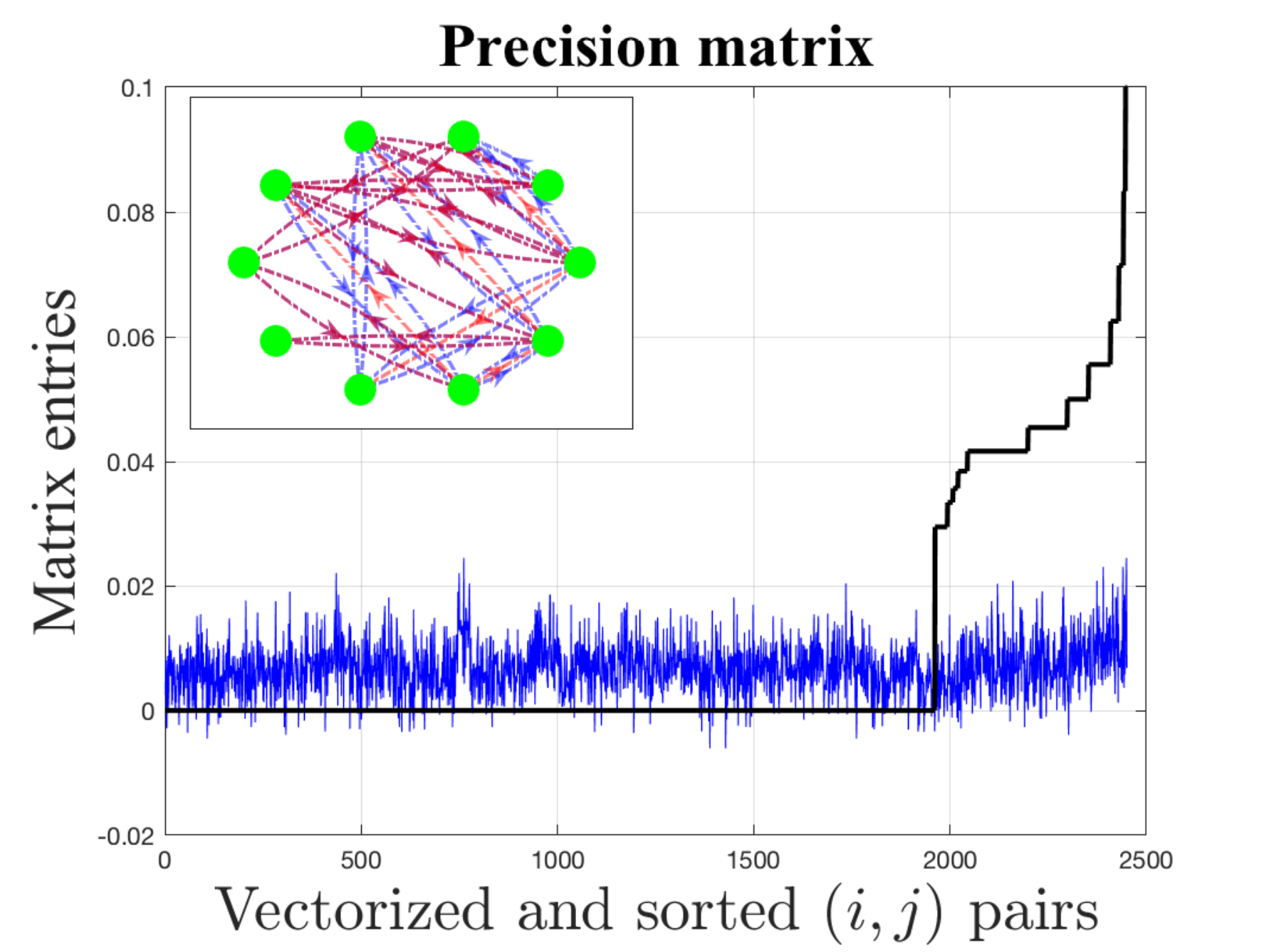}&
\hspace*{-7mm}
\includegraphics[width=44mm]{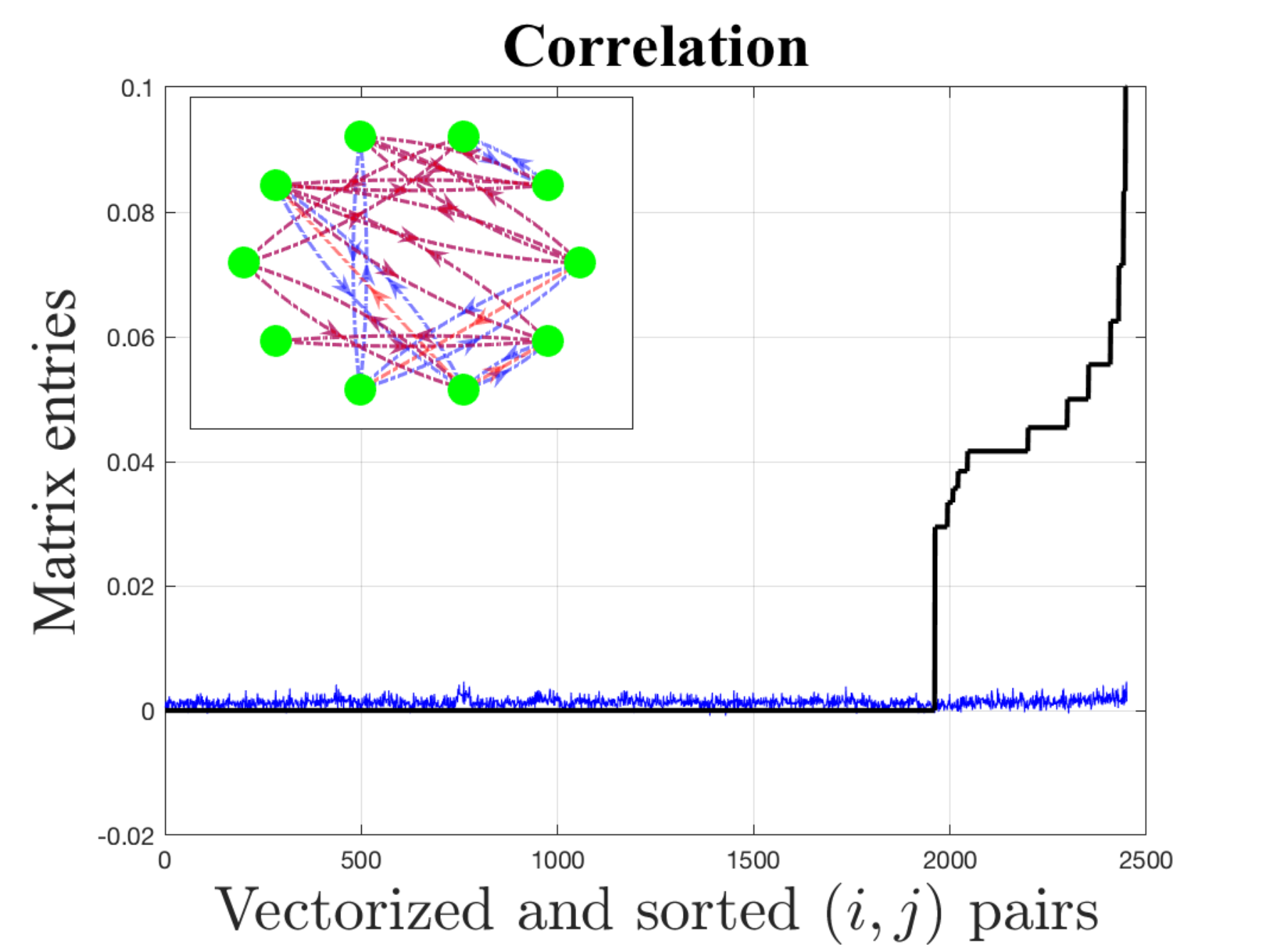}
\end{array}
\]
\caption{Nonlinear dynamical system described by the triple of functions in~(\ref{eq:NLfunExample1sigma})--(\ref{eq:NLfunExample1h}), and depicted in the first three uppermost panels. The true graph is generated according to a binomial graph with connection probability $p=0.2$. The combination matrix follows the uniform averaging rule in~(\ref{eq:unifave}), and is displayed in the fourth uppermost panel. The lowermost panels display the performance of the different estimators as indicated in the panel titles. In the pertinent panels, the matrix entries are vectorized and sorted as described in the main text. 
The inset plot of the uppermost/rightmost panel displays the true graph corresponding to a network portion of $10$ nodes. 
The inset plots of the lowermost panels display the sub-graph learned by the pertinent algorithms, with the red arrows representing edges that do not exist in the true topology, and that are erroneously detected by the learning algorithm.}
\label{fig:Example1}
\end{figure*}

\begin{figure*}[t]
\centering
\[
\begin{array}{cccc}
\hspace*{-5mm}
\includegraphics[width=38.5mm]{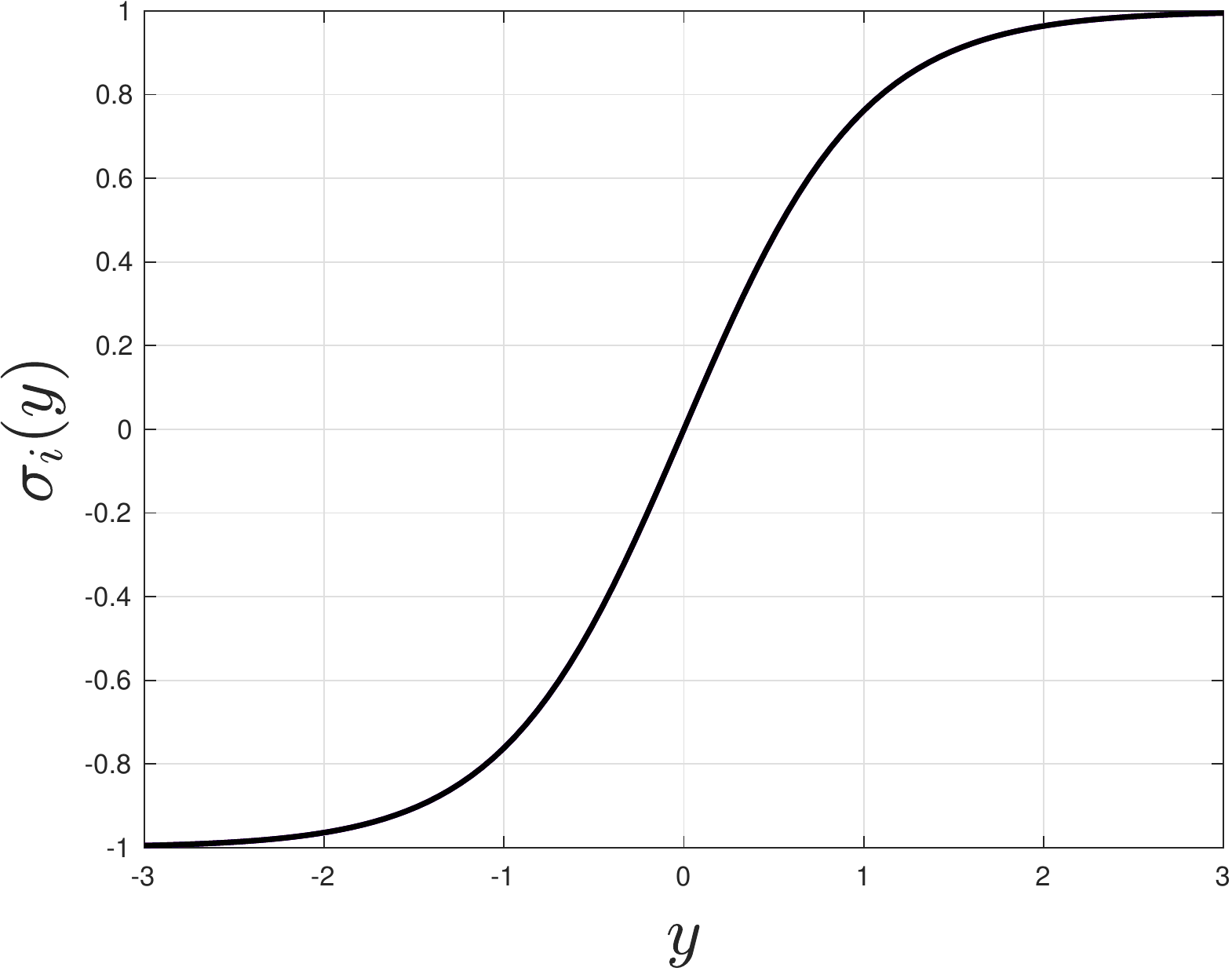}&
\hspace*{-10mm}
\includegraphics[width=38.5mm]{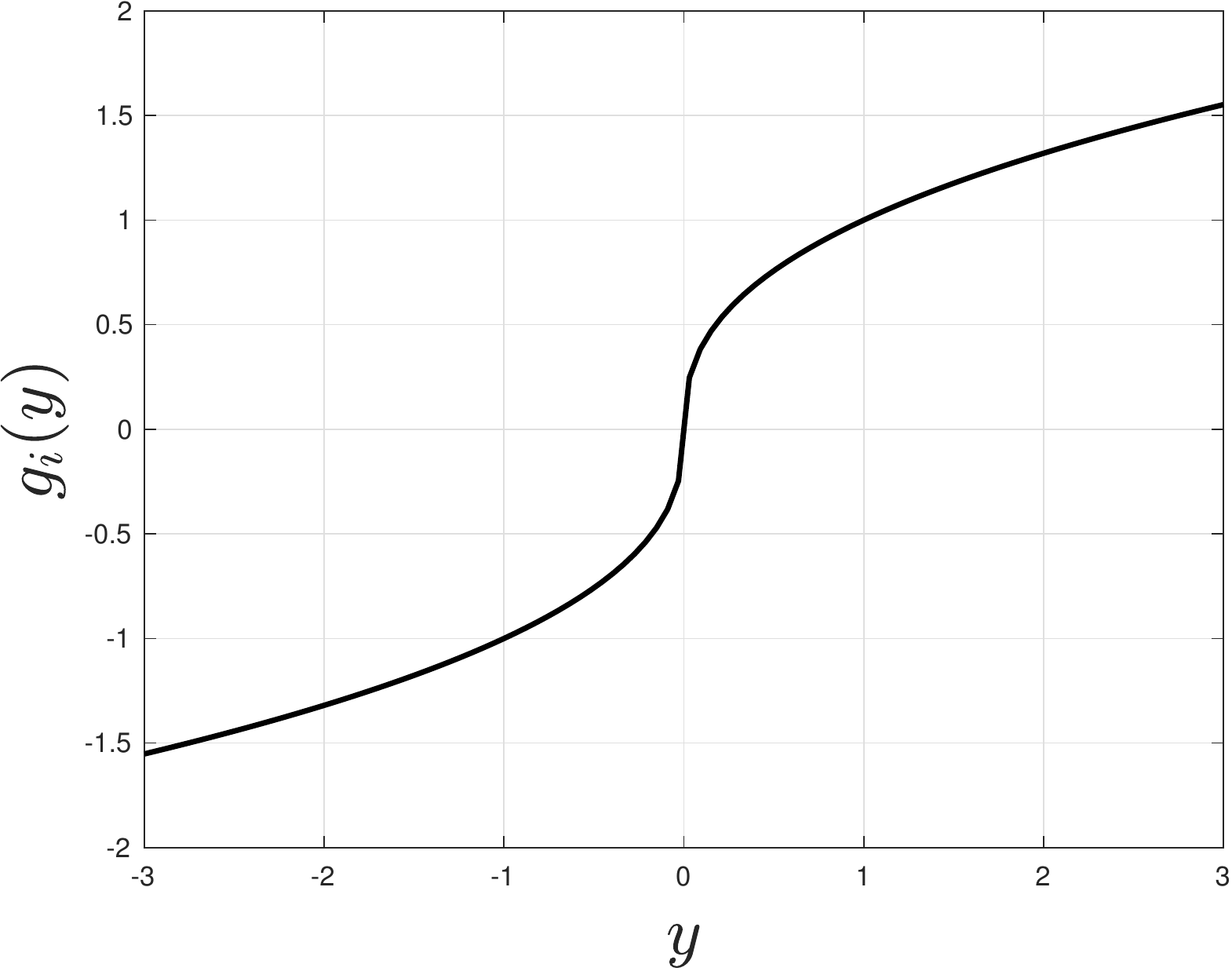}&
\hspace*{-10mm}
\includegraphics[width=38mm]{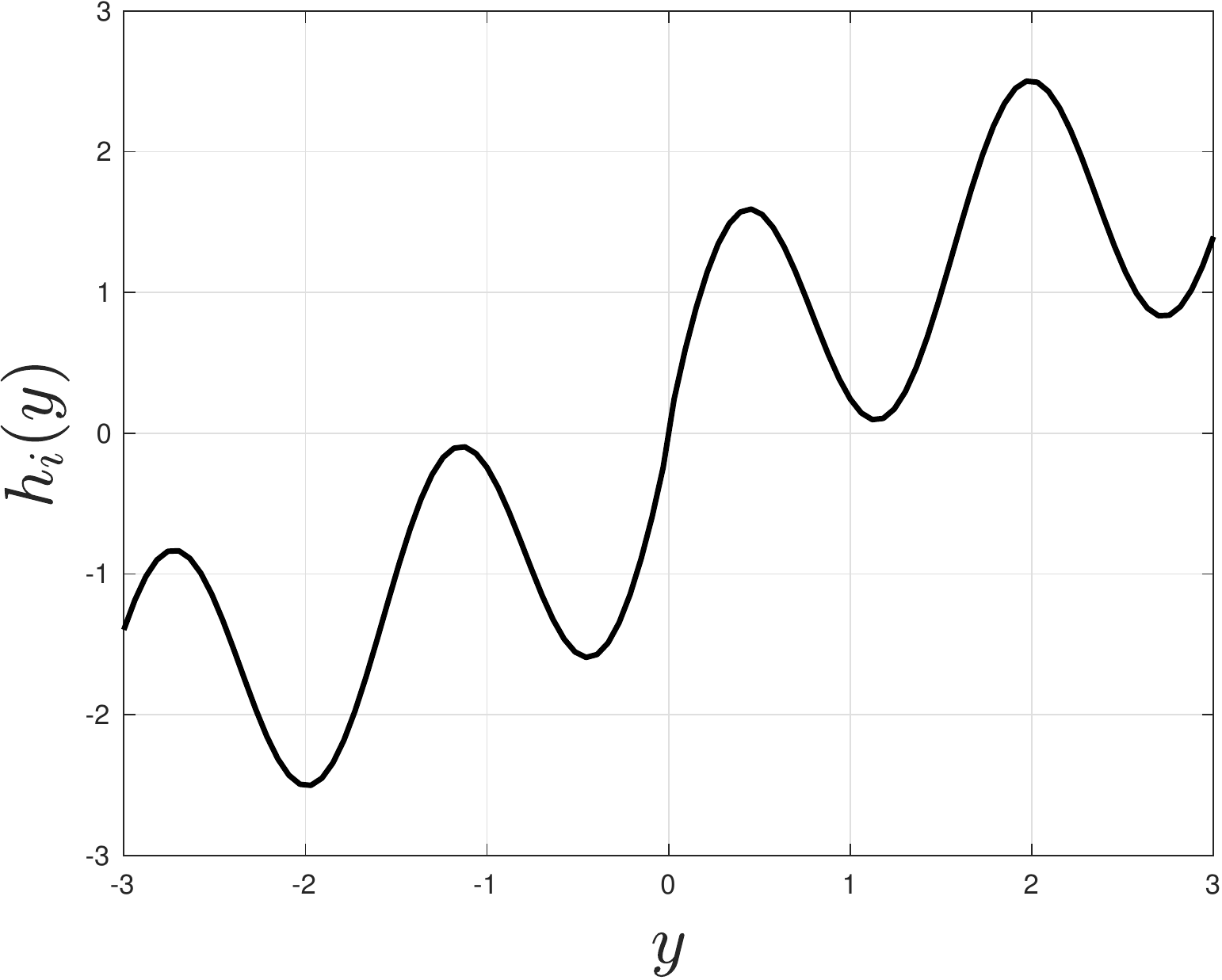}&
\hspace*{-8mm}
\includegraphics[width=44mm]{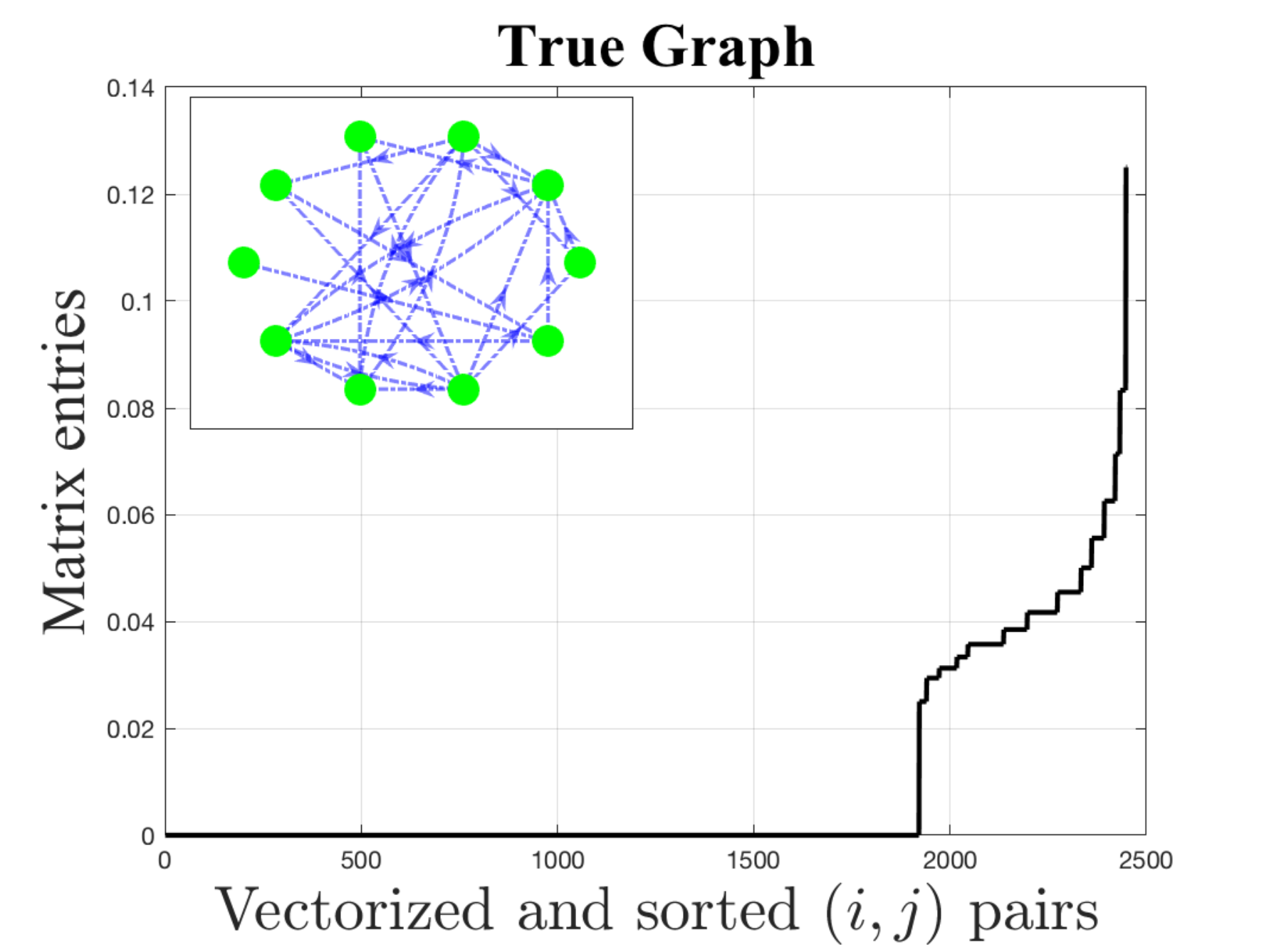}\\
\hspace*{-2mm}
\includegraphics[width=44mm]{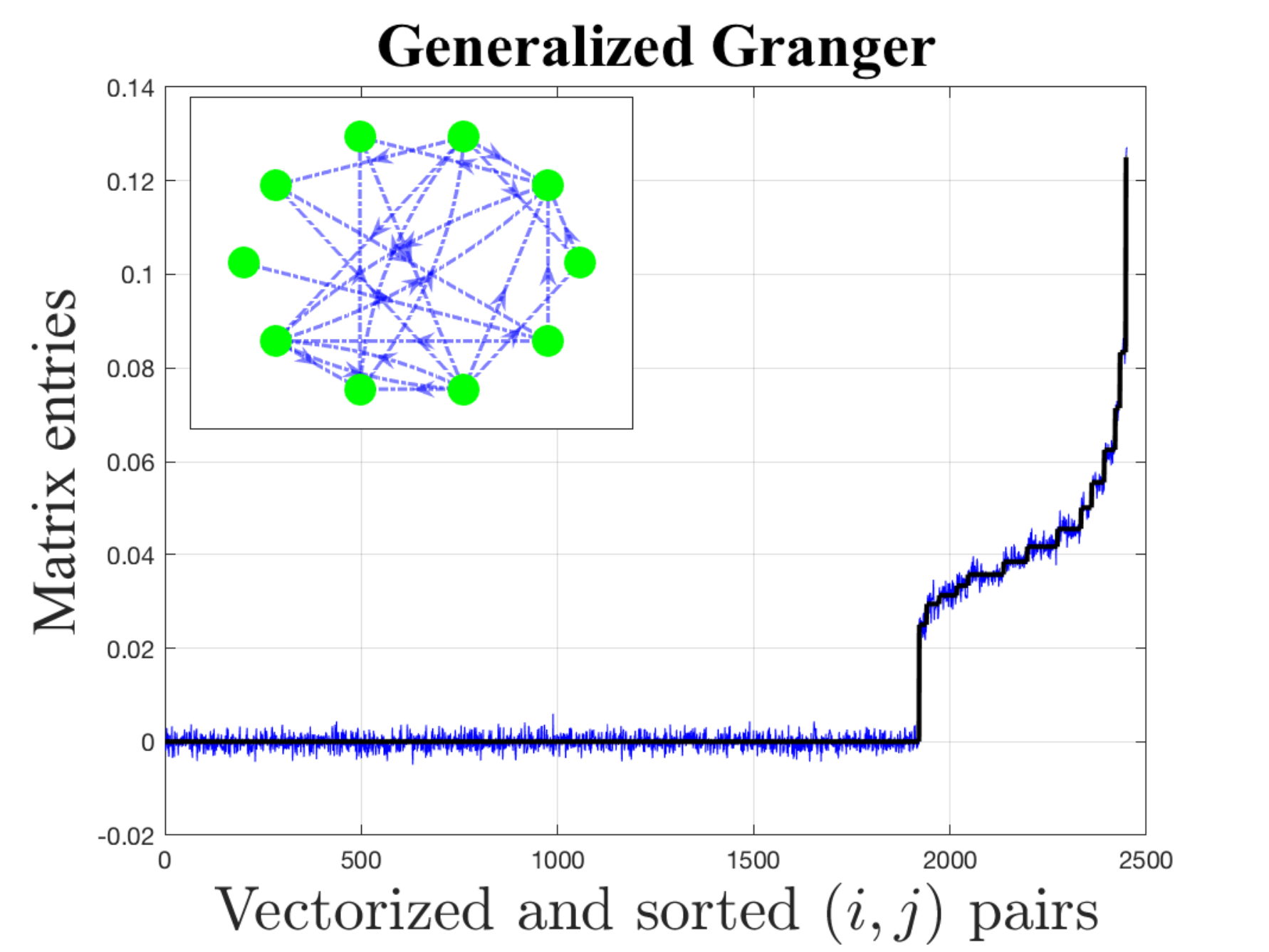}&
\hspace*{-7mm}
\includegraphics[width=44mm]{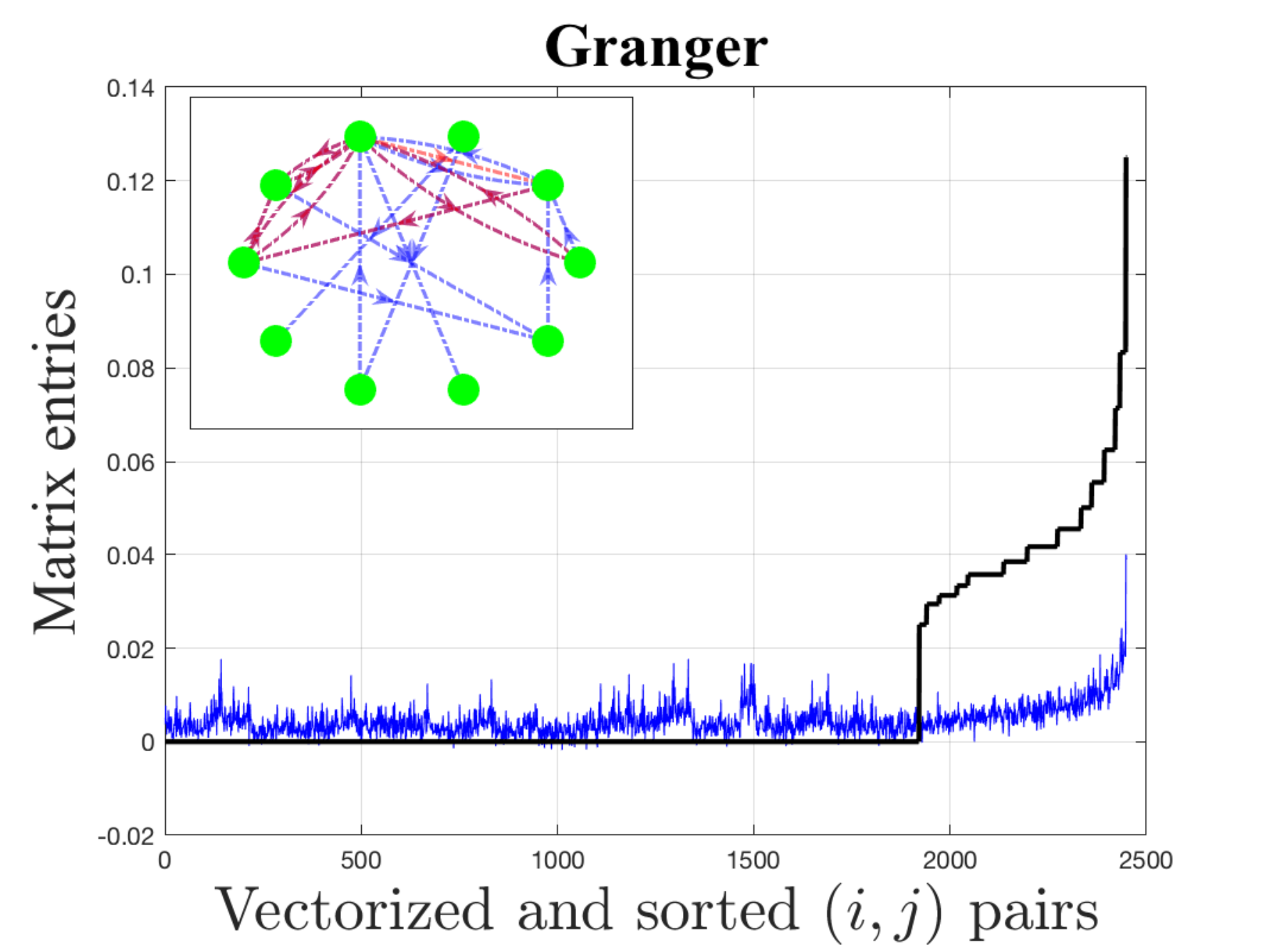}&
\hspace*{-7mm}
\includegraphics[width=44mm]{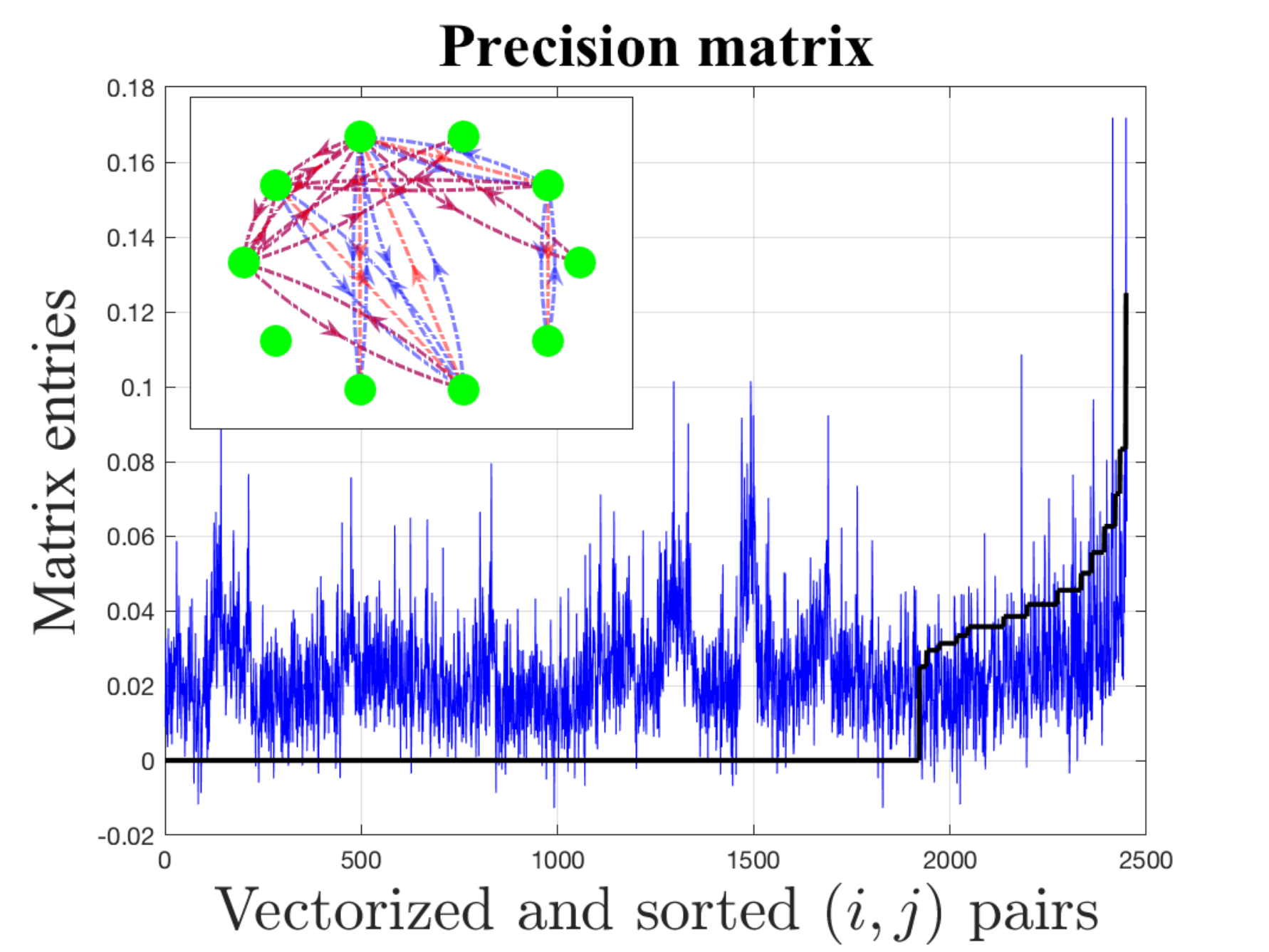}&
\hspace*{-7mm}
\includegraphics[width=44mm]{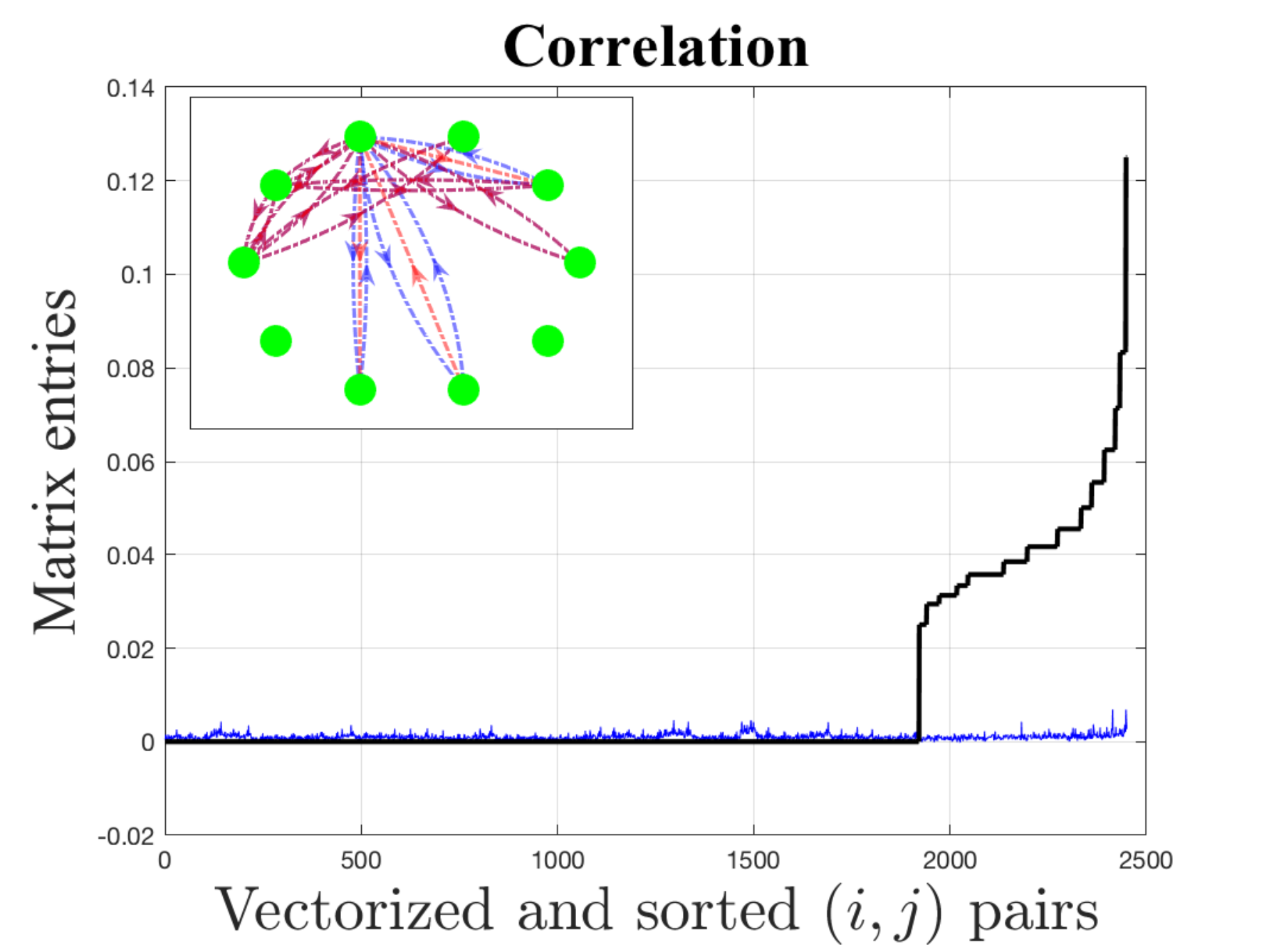}
\end{array}
\]
\caption{Nonlinear dynamical system described by the triple of functions in~(\ref{eq:NLfunExample2sigma})--(\ref{eq:NLfunExample2h}), and depicted in the first three uppermost panels. The true graph is generated according to a binomial graph with connection probability $p=0.2$. The combination matrix follows the uniform averaging rule in~(\ref{eq:unifave}), and is displayed in the fourth uppermost panel. The lowermost panels display the performance of the different estimators as indicated in the panel titles. In the pertinent panels, the matrix entries are vectorized and sorted as described in the main text. 
The inset plot of the uppermost/rightmost panel displays the true graph corresponding to a network portion of $10$ nodes. 
The inset plots of the lowermost panels display the sub-graph learned by the pertinent algorithms, with the red arrows representing edges that do not exist in the true topology, and that are erroneously detected by the learning algorithm.}
\label{fig:Example2}
\end{figure*}

In the following experiments, we will consider a network of $N=50$ nodes. 
We start by examining the full observation case. 
In Fig.~\ref{fig:Example1}, we consider the following nonlinearities to drive our dynamical system, for all $i=1,2,\ldots,N$:
\beqa
\sigma_i(y)&=&{\sf sign}(y)|y|^{0.5},\label{eq:NLfunExample1sigma}\\
g_i(y)&=&{\sf sign}(y)|y|^{0.3},\label{eq:NLfunExample1g}\\
h_i(y)&=&{\sf sign}(y)|y|^{0.7}.\label{eq:NLfunExample1h}
\eeqa
The parameter $\rho$ of the combination matrix is set equal to $0.5$.
In the lowermost row of Fig.~\ref{fig:Example1}, we display the Empirical Generalized Granger (EGG) algorithm and contrast its performance against three standard estimators, namely, the (linear) Granger estimator, the precision matrix (i.e., the inverse of the correlation matrix) and the correlation matrix.
For the sake of a neater data-visualization, and in order to display the matrices values in a one-dimensional plot, we proceeded as follows. First, in all panels the {\em true} matrix $A$ is represented in black. In particular, we first vectorized the {\em true} matrix $A$ and removed the entries associated to its diagonal. As a result, each element in the abscissa of each plot, say $10$, corresponds to a particular entry index, say $(2,6)$. Then, we sorted the entries of the resulting vector in ascending order --- that is why the black curves are nondecreasing. 
Still in the lowermost row of Fig.~\ref{fig:Example1}, the blue curves are obtained by arranging the off-diagonal entries of the pertinent matrix estimators as induced by the aforementioned ordering of $A$. 
In this way, we are contrasting entry-by-entry the ground-truth matrix $A$ with the matrix-estimators in a one-dimensional ordered plot. 
The inset plots displayed in the various panels show just one portion (for the sake of clarity) of the whole network graph learned by the pertinent algorithm. We remark that in this analysis, the algorithm has access to the full network, whereas the panel is limited to a sub-graph just for a matter of visualization.   
In the inset plots, the red arrows represent edges erroneously detected by the learning algorithm (i.e., edges that are not present in the true graph).

Three major observations arise by inspection of Fig.~\ref{fig:Example1}. 
First, we see that the EGG algorithm is able to estimate faithfully the combination weights. 
Second, the clustering algorithm is able to properly reconstruct the network skeleton from the estimated combination matrix.
Third, and perhaps not unexpectedly, the classical methods that work in the linear case (Granger), or in the Gaussian graphical model setting (precision), or over correlation-networks (correlation matrix), are essentially blind in our nonlinear framework.

In Fig.~\ref{fig:Example2}, we repeat the experiment on another set of nonlinearities, namely,
\beqa
\sigma_i(y)&=&\tanh{y},\label{eq:NLfunExample2sigma}\\
g_i(y)&=&{\sf sign}(y)|y|^{0.4},\label{eq:NLfunExample2g}\\
h_i(y)&=&\sin(4 y) + {\sf sign}(y)|y|^{0.6}.\label{eq:NLfunExample2h}
\eeqa 
We see that similar conclusions apply. One interesting difference is the better convergence of the EGG and of the one-lag-functional estimators, which could be ascribed to the fact that the activation function, $\sigma$, is now bounded, a feature that concurs to increase the system stability, and, hence, the speed of convergence of the various empirical estimators.

\begin{figure}[t]
\centering
\includegraphics[width=80mm]{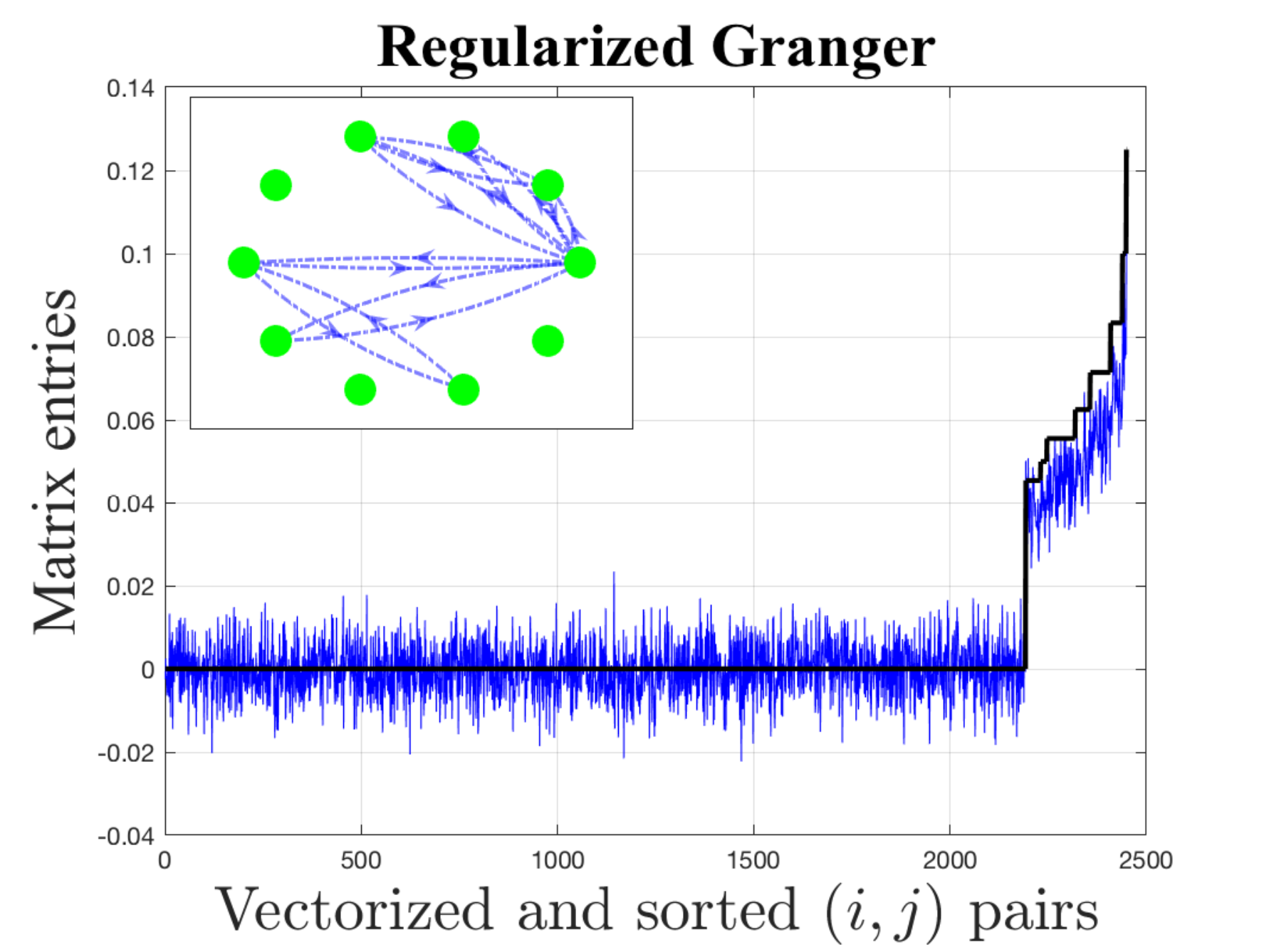}
\includegraphics[width=80mm]{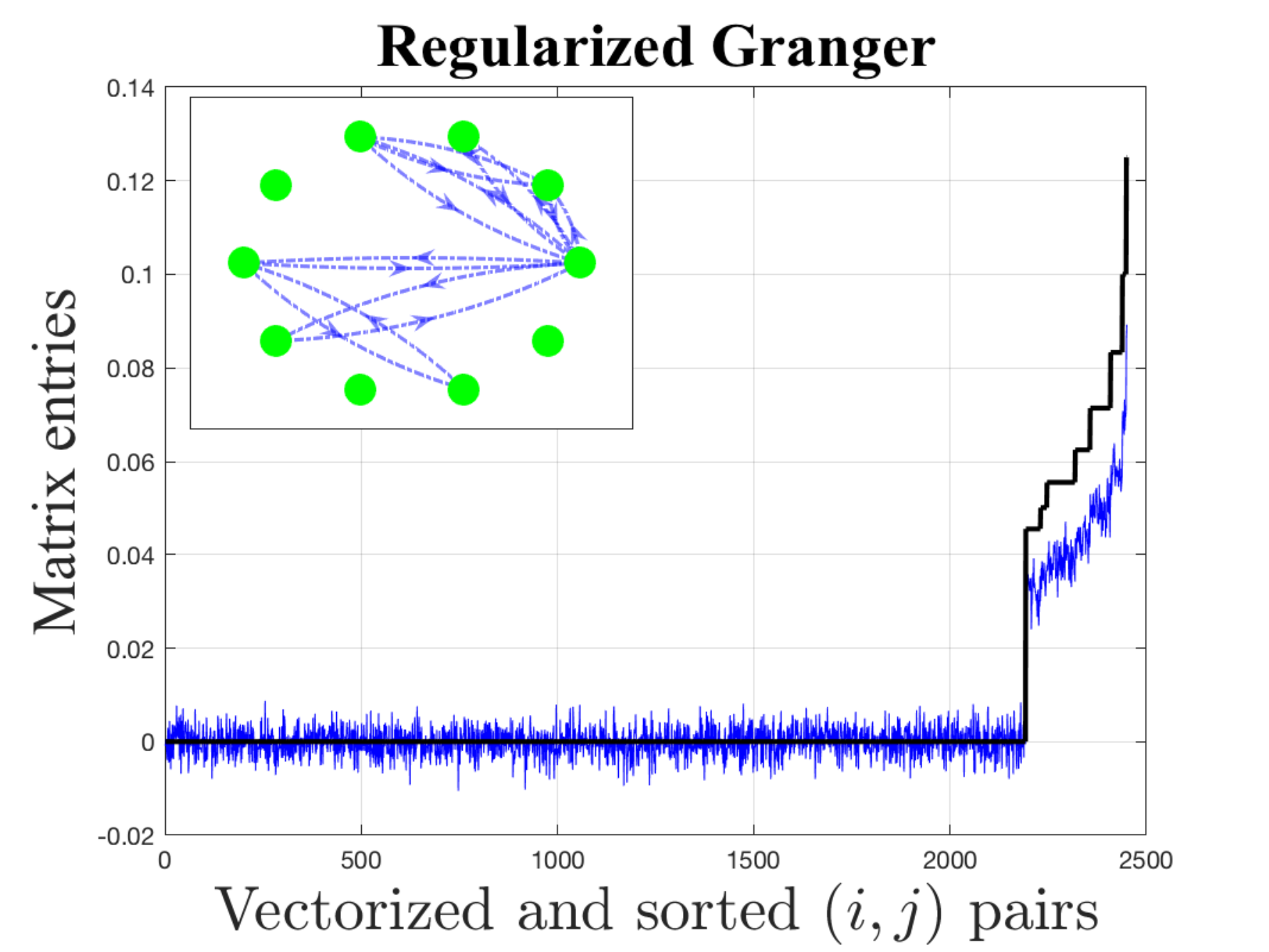}
\caption{Same example as in Fig.~\ref{fig:h1sing}, but with a {\em regularized} weighting function $\widetilde{\omega}(y)$. In the leftmost panel, $\widetilde{\omega}(y)$ is equal to $\omega(y)$ outside the interval $(-0.1,0.1)$. In the rightmost panel, a stronger perturbation is applied to $\omega(y)$, since $\widetilde{\omega}(y)$ is equal to $\omega(y)$ outside the interval $(-0.2,0.2)$.}
\label{fig:h1regul}
\end{figure}

\section{Beyond the Theoretical Results}
\label{sec:partial}
The results summarized in Proposition~\ref{theor:ergodicity} allow consistent estimation of the network graph under the specific setting and under a set of assumptions that we have extensively discussed. 
On the other hand, there exist relevant scenarios where the setting need to be enlarged and some of the assumptions relaxed. 
In this section, we show how the analysis can be helpful to address some of these scenarios.

\subsection{Regularization of the Weighting Function $\omega(y)$}
From our analysis, it is apparent that one limitation is the integrability condition~(\ref{eq:oneoverh1boundK}) imposed on the weighting function $\omega(y)$. For example, in Fig.~\ref{fig:h1sing} we have seen that a linear $g(y)$ can lead to a weighting function $\omega(y)$ with infinite expectation (because of the too-fast singularity of $g(y)=y$ around the origin), resulting in a singular estimator. 
In order to remedy this issue, one could replace $\omega(y)$ with a regularized version, $\widetilde{\omega}(y)$, in the evaluation of the one-lag function $F_1(\bm{y}_{n+1},\bm{y}_{n})$. For example, we could set $\widetilde{\omega}(y)=\omega(y)$ outside some small neighborhoods of its singularities, and $\widetilde{\omega}(y)$ equal to some constant within these neighborhoods. 
The rationale behind this regularization is that, on one hand, we rule out the pathological behavior of the estimator since we remove the singularities; on the other hand, for sufficiently small perturbations of the original $\omega(y)$, we expect that the deviations from the true combination matrices are small enough to let the clustering algorithm be still able to classify correctly the links between nodes.

We apply the proposed regularization to the example of Fig.~\ref{fig:h1sing}, and the result is shown in Fig.~\ref{fig:h1regul}. 
Some notable features emerge. 
We start by examining the leftmost panel. 
First, we see that the regularized weighting function removes the singular behavior of the estimator (blue curve), which is now capable to follow the true profile (black curve) of the combination matrix entries. 
This behavior is reasonable because our regularization has in fact removed the singularity.

Second, we observe that the estimator looks noisier as compared to the examples of Figs.~\ref{fig:Example1} and~\ref{fig:Example2}. 
This augmented irregularity can be explained because we started from a {\em singular} weighting function.
Third, we see that the estimator seems not to converge to the true combination matrix, which is expected because consistent reproduction of the combination matrix is not granted when we use a surrogate weighting function.
More specifically, a trade-off arises between the degree of irregularity and the fidelity of reconstruction, and we expect that the smaller the perturbation of the original weighting function is, the higher the irregularity of the estimator and the fidelity of reconstruction will be. 
This trade-off is confirmed in the rightmost panel of Fig.~\ref{fig:h1regul}, where we consider a stronger perturbation (i.e., we modify the weighting function in an ampler neighborhood). Comparing the rightmost panel against the leftmost panel, we see that the blue curve is now less wild, but that it is more distant from the true matrix (black curve). 
In summary, a sort of uncertainty principle is exhibited: one can either get a more precise knowledge (less oscillating curves) of a less precise matrix value; or a less precise knowledge (i.e., more oscillating curves) of a matrix value closer to the true value.

This notwithstanding, we should keep in mind that the basic goal of graph learning is retrieving the skeleton of the network, i.e., the support graph of the combination matrix. For this reason, it is not so crucial that the estimator is not able to reproduce exactly the values of the combination matrix. What is critical is that the {\em identifiability gap} between the estimated matrix entries corresponding to disconnected or connected node pairs is still well recognizable from the estimator, which would allow to the clustering algorithm to recover the correct graph. The inset panels of Fig.~\ref{fig:h1regul} show that this can be in fact possible. 
This notion of identifiability gap has been introduced and extensively discussed in~\cite{MattaSantosSayedAsilomar2018,MattaSantosSayedISIT2019}, and will play a role especially in the partial observation setting, as we will see in the forthcoming section.

\subsection{Partial Observation Setting}
All the analysis conducted so far was based on a full-observability assumption, namely, on the assumption that samples from all nodes can be collected. 
We note however that in complex large-scale systems, we are often unable to probe the state-evolution of {\em all} nodes. 
Accordingly, in this section we consider the partial observation setting where the data can be collected from only a subset $\mathcal{S}$ of nodes. The objective of the learning becomes then reconstructing the support graph of the partial sub-matrix, $A_{\mathcal{S}}$, namely, of the sub-matrix corresponding to subset $\mathcal{S}$. 
Likewise, in the partial observation case we assume that the truncated matrix functions\footnote{We remark that the sub-matrices $[\mathcal{F}_0]_{\mathcal{S}}$ and $[\mathcal{F}_1]_{\mathcal{S}}$ can be constructed from the samples $\{\bm{y}_n\}$ because the $i$-th component of the functions $\sigma(y), h(y), g(y), g(y)$ depends solely on the $i$-th entry $y_i$.}:
\beq
[\mathcal{F}_0]_{\mathcal{S}},~~[\mathcal{F}_1]_{\mathcal{S}},
\eeq
will replace the full matrices $\mathcal{F}_0$ and $\mathcal{F}_1$.

However, since $A=\mathcal{F}_1\mathcal{F}_0^{-1}$, in general the latent (unobserved) samples for nodes outside $\mathcal{S}$ affect the possibility of constructing $A_{\mathcal{S}}$ from $[\mathcal{F}_0]_{\mathcal{S}}$ and $[\mathcal{F}_1]_{\mathcal{S}}$. 
For example, we have:
\beq
A_{\mathcal{S}}\neq [\mathcal{F}_1]_{\mathcal{S}} ([\mathcal{F}_0]_{\mathcal{S}})^{-1}.
\label{eq:partialGGE}
\eeq
For the special case of {\em linear} networked dynamical systems, the generalized Granger boils down to the Granger estimator, and the partial construction in~(\ref{eq:partialGGE}) is obtained by considering the zero-lag and one-lag correlation matrices. 
Recent works have in fact established the structural consistency of this partial (i.e., applied only to a subset of nodes) Granger estimator in the linear case, for a class of regular symmetric combination matrices $A$ (that include, for instance the classic Laplacian and Metropolis matrices), and when the underlying graph is an undirected Erd\H{o}s-R\'enyi graph under various regimes of connectivity~\cite{tomo_icassp,tomo,tomo_dsw,tomo_isit,MattaSantosSayedAsilomar2018,MattaSantosSayedISIT2019}.
In particular, in~\cite{tomo} feasibility of such inverse problem is proved for sparse Erd\H{o}s-R\'enyi graphs and increasing cardinality of the observable space; the analysis is extended in~\cite{tomo_dsw,tomo_isit} to cover the relevant case where the observed nodes have {\em arbitrary connection structure} and the cardinality of observed nodes is fixed (thus, the degree of observability is low); in~\cite{MattaSantosSayedAsilomar2018,MattaSantosSayedISIT2019}, the analysis is extended to cover the case of densely connected networks. 

\begin{figure*}[t]
\centering
\[
\begin{array}{cc}
\includegraphics[width=71mm]{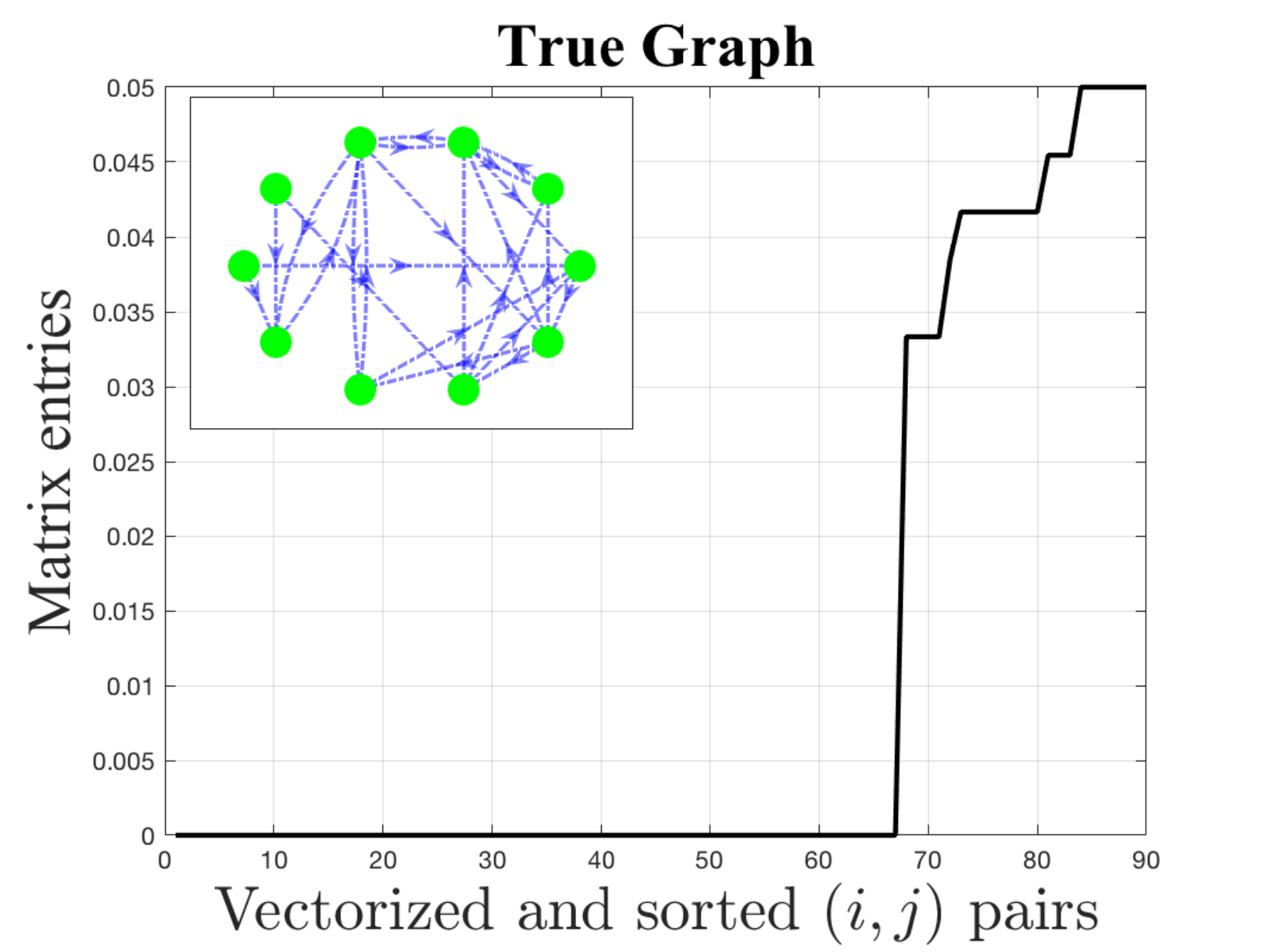}
\includegraphics[width=71mm]{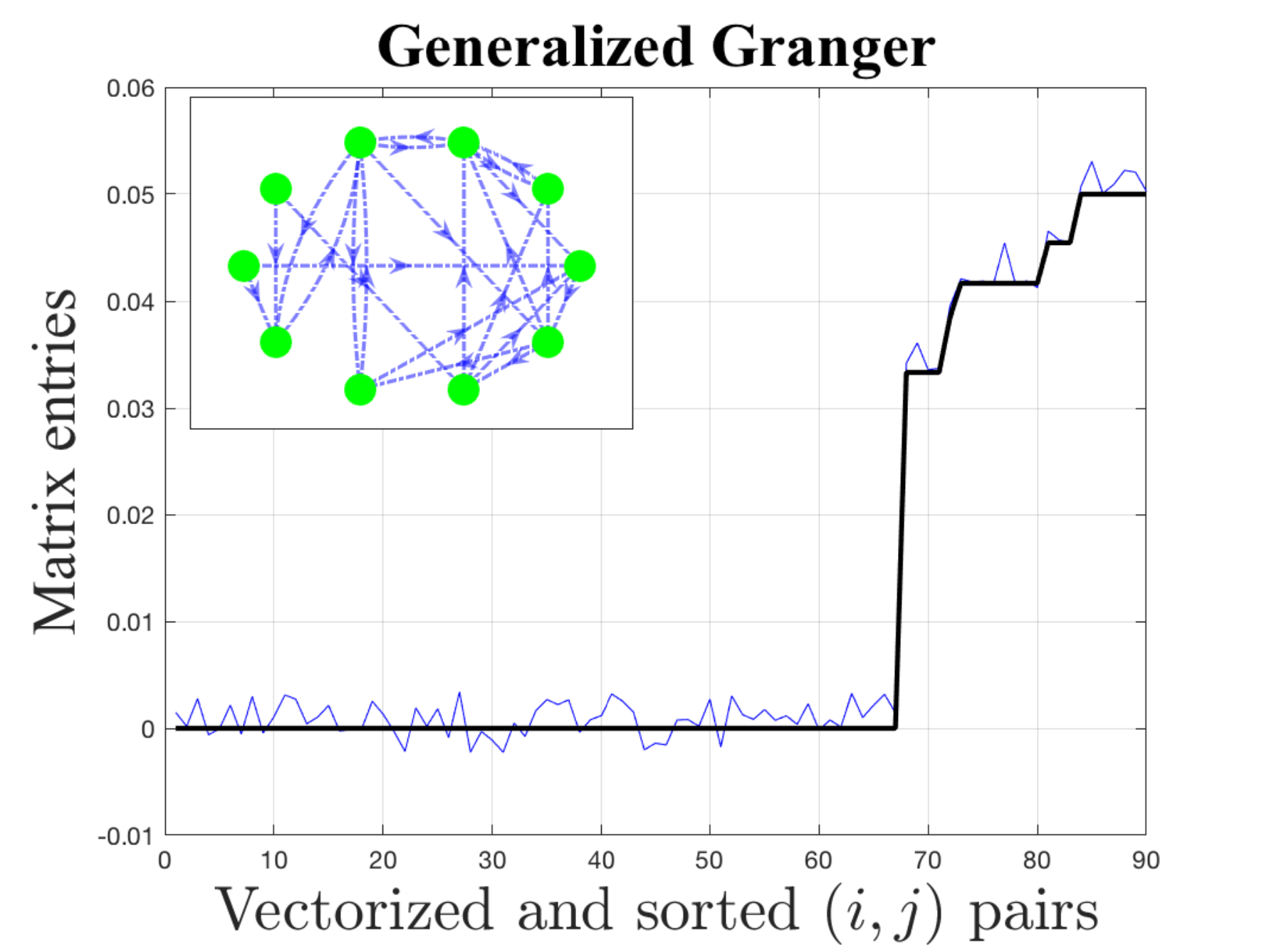}\\
\includegraphics[width=71mm]{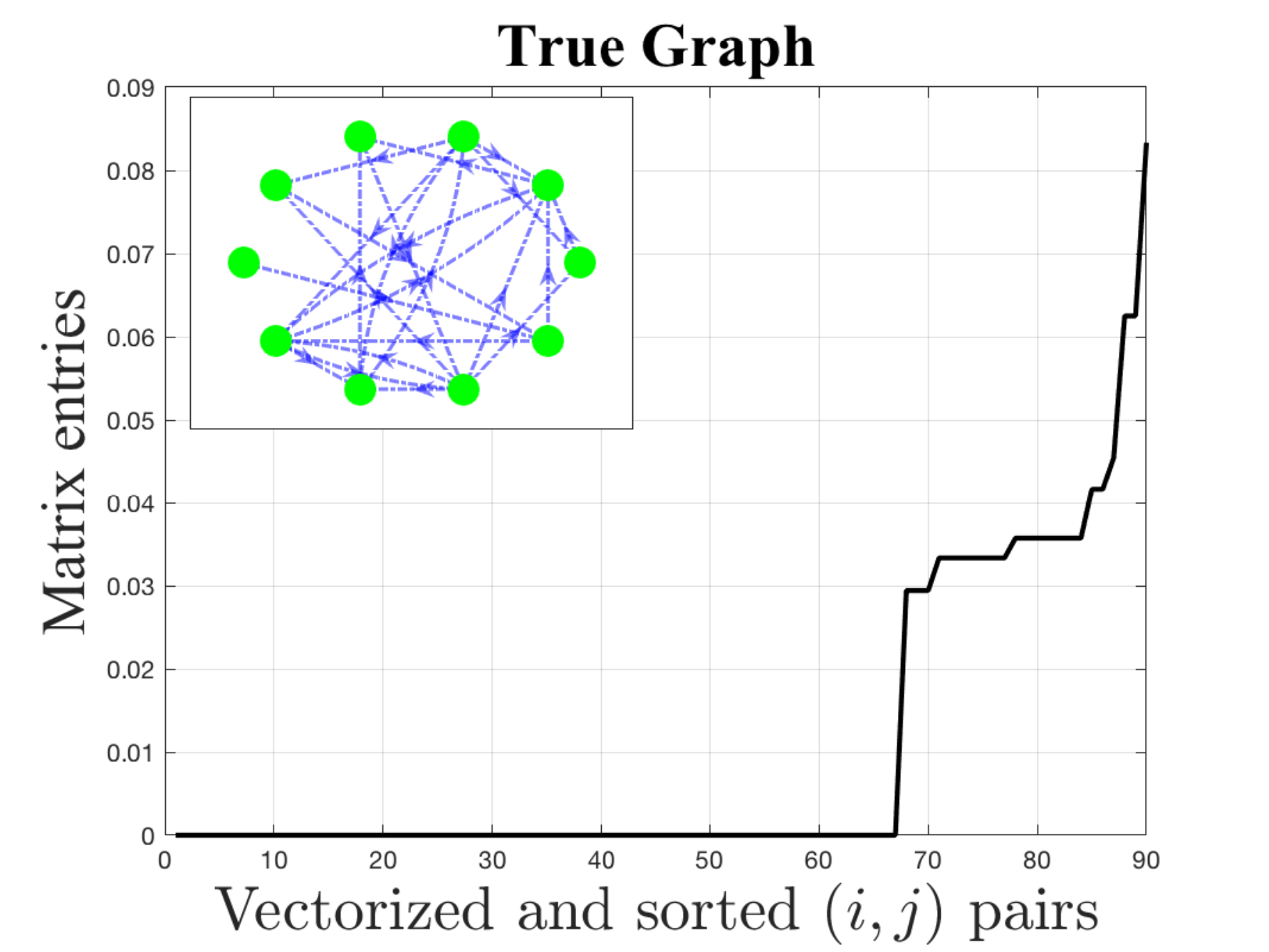}
\includegraphics[width=71mm]{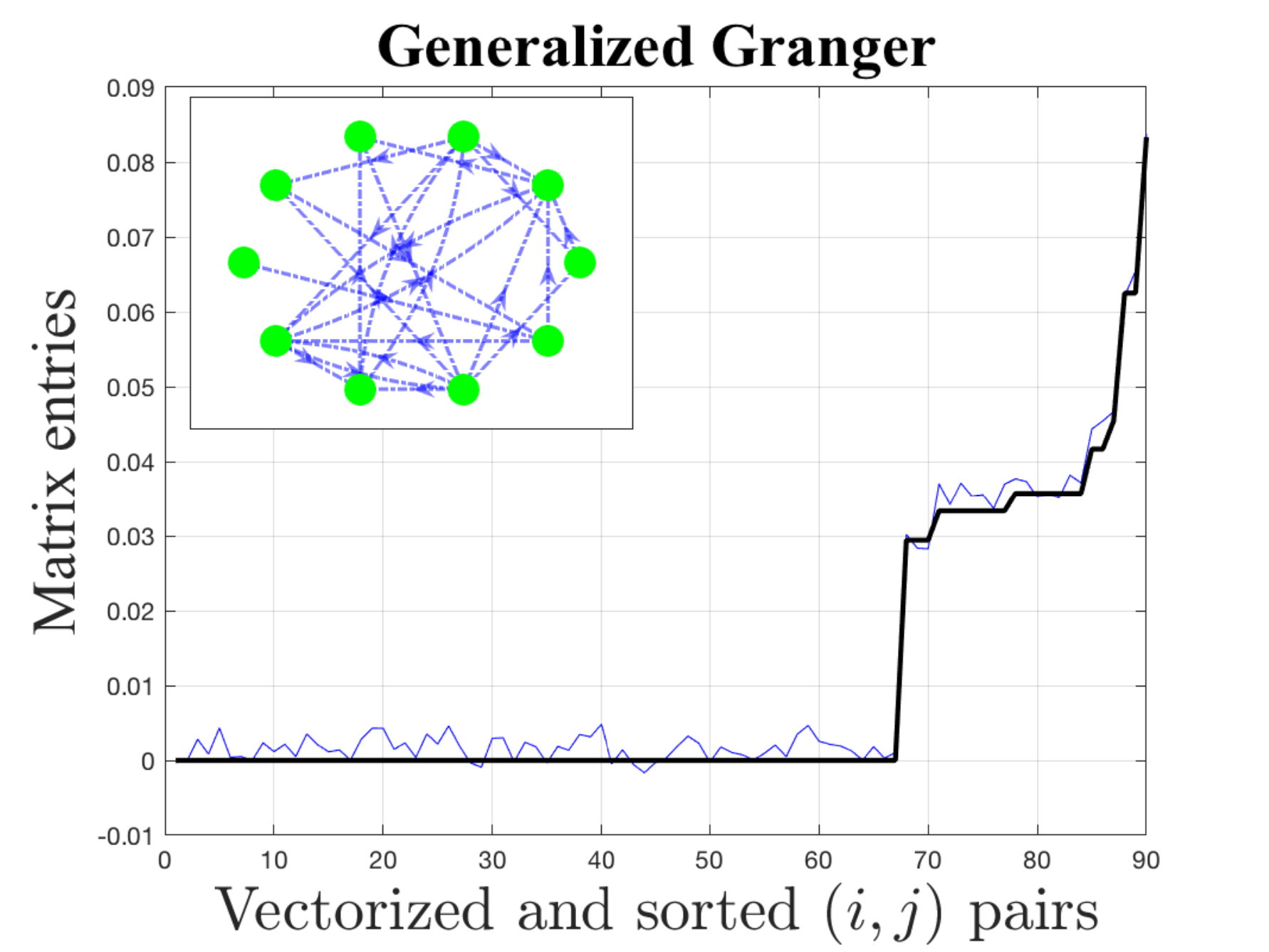}
\end{array}
\]
\caption{
{\em Partial} generalized Granger estimator in~(\ref{eq:partialGGE}) applied when only $10$ out of $50$ nodes are probed. The inset plots display the sub-graph of probed nodes as reconstructed by the learning algorithm.
{\em Uppermost panels}: Same example as in Fig.~\ref{fig:Example1}. 
{\em Lowermost panels}: Same example as in Fig.~\ref{fig:Example2}. 
}
\label{fig:Example12Partial}
\end{figure*}

Motivated by these results holding for the linear case, we now test the partial Generalized Granger estimator in~(\ref{eq:partialGGE}) with the more general nonlinear dynamical systems addressed in this work. 
Devising technical guarantees for structural consistency of the generalized Granger under this latent scenario appears to be a highly nontrivial task.
Therefore, we are not in the position here to prove that the proposed method is consistent under this particular latent regime. 
Nevertheless, it is useful to see whether, under the conditions used to examine the nonlinear model under full-observation, the algorithm applied to partial observation can work.

In Fig.~\ref{fig:Example12Partial}, we use the same settings adopted in Figs.~\ref{fig:Example1} and~\ref{fig:Example2}, but for one essential detail. 
Now, only a subset $\mathcal{S}$ with $10$ out of $50$ nodes is accessible. According to the discussion in Sec.~\ref{sec:partial}, we implement two estimators, namely, the EGG with $[\widehat{F}_0(n)]_{\mathcal{S}}$ and $[\widehat{F}_1(n)]_{\mathcal{S}}$, namely, with the matrix functions evaluated {\em only on data coming from the observed network component}.
We see how both the proposed method is able to faithfully estimate the network graph. 

We remark that the proposed analysis is by no means exhaustive, and is meant to show that there are cases where the generalized Granger method can work in the nonlinear regime. 
On the other hand, even if we do not have a definite answer, we have also evidences that it can fail, and that it is in particular sensitive to two features: $i)$ heterogeneity, namely, when the nonlinear components at different nodes behave very differently, the role of some latent nodes might become dominant and corrupt the fidelity of the graph reconstruction; $ii)$ level of noise, which is a distinguishing feature of the nonlinear setting, since in the linear case the noise variance played basically as a scale factor that does not affect the graph identifiability. Perhaps not unexpectedly, in the nonlinear case the level of noise can alter even substantially the overall qualitative behavior of the dynamics.

\appendices

\section{Proof of Proposition~\ref{theor:integrability}}
\label{app:integrability}

\begin{lemma}
\label{lem:oneoverh1}
Under Assumption~\ref{assum:oneoverh1}, we have, for all $n\in\mathbb{N}$:
\beq
\E[\|\omega(\bm{y}_n)\|^2]\leq K.
\eeq
\end{lemma}

\begin{IEEEproof}
Let $k(y)=g(y)\odot A h(y)$. In view of~(\ref{eq:ultimodelvec}), we have that $\bm{y}_n=\sigma(k(\bm{y}_{n-1}) + \bm{x}_n)$. 
We can use the tower property to write:
\beq
\E\left[\|\omega(\bm{y}_n)\|^2\right]
=
\E\left[
\,\E\left[\|\omega(\sigma(k(\bm{y}_{n-1}) + \bm{x}_n))\|^2\,\,|\bm{y}_{n-1}\right]
\,
\right]
\eeq
However, since $\bm{x}_n$ is statistically independent from $\bm{y}_{n-1}$, we have:
\beqa
\lefteqn{\E\left[\|\omega(\sigma(k(\bm{y}_{n-1}) + \bm{x}_n))\|^2\,\,|\bm{y}_{n-1}\right]}\nonumber\\
&\leq&
\sup_{c\in\mathbb{R}}
\E\left[\|\omega(\sigma(c + \bm{x}))\|^2\right]\leq K,
\label{eq:wboundineq}
\eeqa
where in the first inequality we used the fact that whenever $\bm{x}$ and $\bm{y}$ are independent random variables, and for any measurable map $q$, the underlying function $r$
\beq
r(\bm{y})\dfz\E\left[q(\bm{x},\bm{y})\left|\bm{y}\right.\right]
\eeq
is given by $r(y)=\mathbb{E}\left[q(\bm{x},y)\right]$ for all $y\in\mathbb{R}^N$~\cite{kallenberg2002} and we have that $r(\bm{y})\leq \sup_{c\in\mathbb{R}^N}r(c)$ almost surely (which yields the first inequality in~(\ref{eq:wboundineq}) with the proper choice of $q$). 
The latter inequality in~(\ref{eq:wboundineq}) holds in view of~(\ref{eq:oneoverh1boundK}). 
\end{IEEEproof}

\begin{remark}
Note that the integrability condition on $\omega(\bm{y}_n)$ implies that the vector $g(\bm{y}_n)$ has some zero entry with zero probability.~\hfill$\square$ 
\end{remark}

\begin{lemma}
\label{lem:ynormbound}
Let us define, for an arbitrarily small $\epsilon>0$, the following constant:
\beq
\gamma\dfz\left\{
\begin{array}{ll}
\alpha_{\sigma}\alpha_g\alpha_h\|A\|+\epsilon,~~\textnormal{if $p>0$ and $q>0$}\\
\alpha_{\sigma}\alpha_g\beta_h\|A\|,~~~~~~~\textnormal{if $p=1$ and $q=0$}\\
\alpha_{\sigma}\alpha_h\beta_g\|A\|,~~~~~~~\textnormal{if $p=0$ and $q=1$}
\end{array}
\right.
\label{eq:gammadef}
\eeq
where $p$, $q$ and the various $\alpha$ and $\beta$ constants have been introduced in~(\ref{eq:newLipschitzlikesigma})--(\ref{eq:newLipschitzlikeh0}). 
We have that:
\beq
\|h(\bm{y}_n)\|\leq \alpha_h
\left(
\gamma^n \|\bm{y}_0\| + \alpha_{\sigma}\,\sum_{i=1}^n \,\gamma^i \|\bm{x}_{n-i}\|
\right)
+\kappa,
\eeq
for some $\kappa>0$.
\end{lemma}

\begin{IEEEproof}
Using~(\ref{eq:newLipschitzlikesigma})--(\ref{eq:newLipschitzlikeh0}) we have the following chain of inequalities:
\beqa
\lefteqn{\| \sigma\left( g(y)\odot  A h(y) + x\right) \|} 
\nonumber\\
&\leq& 
\alpha_{\sigma} \|g(y)\odot A h(y) + x\| + \beta_{\sigma}\nonumber\\
&\leq&
\alpha_{\sigma} (\|g(y)\odot A h(y)\|+\|x\|) + \beta_{\sigma}\nonumber\\
&\leq&
\alpha_{\sigma}  \|A\| \, \|D_{g}(y)\| \, \|h(y))\|  
+
\alpha_{\sigma}\|x\| + \beta_{\sigma}\nonumber\\
&\leq&
\alpha_{\sigma} \|A\|
\left(
\alpha_g\|y\|^p + \beta_g
\right)
\left(
\alpha_h\|y\|^q + \beta_h
\right)
+
\alpha_{\sigma}\|x\| + \beta_{\sigma}
\nonumber\\
&\leq&
\alpha_{\sigma} \alpha_g\alpha_h \|A\| \|y\|^{(p+q)} + 
\alpha_{\sigma}\|x\| + \beta_{\sigma}+\alpha_{\sigma} \|A\|\beta_g\beta_h
\nonumber\\
&+&
\alpha_{\sigma} \alpha_g\beta_h\|A\|  \|y\|^p
+
\alpha_{\sigma} \alpha_h\beta_g\|A\| \|y\|^q.
\label{eq:chaineq}
\eeqa
Now, consider first the case where both $p>0$ and $q>0$. In this case, since $p+q=1$, we have that both $p$ and $q$ are strictly less than $1$, implying that, for an arbitrarily small $\epsilon$, and for a certain constant $c$ we can write:
\beqa
\alpha_{\sigma} \alpha_g\beta_h\|A\|  \|y\|^p &\leq& c + \frac\epsilon 2\|y\|,\nonumber\\
\alpha_{\sigma} \alpha_h\beta_g\|A\|  \|y\|^p &\leq& c + \frac\epsilon 2\|y\|.
\label{eq:boundingpq}
\eeqa
Accordingly, from~(\ref{eq:chaineq}) we get:
\beq
\| \sigma\left( g(y)\odot  A h(y) + x\right) \|\leq
\gamma \|y\| + \alpha_{\sigma}\|x\| + \beta,
\eeq
where we have collected all the constant terms into the factor $\beta$, and we have used the first definition of $\gamma$ in~(\ref{eq:gammadef}).

Next consider the case $q=0$. This implies that $h(y)$ is bounded by a constant, and we accordingly have $\alpha_h=0$. 
Therefore, Eq.~(\ref{eq:chaineq}) becomes:
\beq
\| \sigma\left( g(y)\odot A h(y) + x\right) \|
\leq
\gamma \|y\| + \alpha_{\sigma}\|x\| + \beta,
\eeq
where we have collected the constant terms into $\beta$, and where we have used the second definition of $\gamma$ in~(\ref{eq:gammadef}). The proof for the case $p=0$ follows similarly.
\end{IEEEproof}

\begin{lemma}
\label{lem:UIh2}
If the constant $\gamma$ in~(\ref{eq:gammadef}) is strictly smaller than $1$, then the sequence of random variables $\{\|h(\bm{y_n})\|^2\}$ is uniformly integrable. 
\end{lemma}

\begin{IEEEproof}
First, we observe that:
\beq
\bm{u}_n\dfz\|h(\bm{y_n})\|^2=\varphi(\bm{y}_0,\bm{x}_1,\bm{x}_2,\ldots,\bm{x}_n),
\eeq
for a deterministic function $\varphi$. 
In view of Lemma~\ref{lem:ynormbound}, we can write:
\beqa
\lefteqn{
\varphi(\bm{y}_0,\bm{x}_1,\bm{x}_2,\ldots,\bm{x}_n)}
\nonumber\\
&\leq&
\left[
\alpha_h
\left(
\gamma^n \|\bm{y}_0\| + \alpha_{\sigma}\,\sum_{i=1}^n \,\gamma^i \|\bm{x}_{n-i}\|
\right)
+\kappa\right]^2.
\label{eq:fbound}
\eeqa
Now, since the $\bm{x}_i$'s are i.i.d., the function:
\beq
\bm{u}'_n=\varphi(\bm{y}_0,\bm{x}_n,\bm{x}_{n-2},\ldots,\bm{x}_1)
\label{eq:zprime}
\eeq
that is obtained by applying the function $\varphi$ to a reversed sequence $\bm{x}_n,\bm{x}_{n-1},\ldots,\bm{x}_1$ {\em has the same distribution} of $\bm{u}_n$. Applying~(\ref{eq:zprime}) to~(\ref{eq:fbound}) we can write:
\beqa
\lefteqn{
\bm{u}'_n=\varphi(\bm{y}_0,\bm{x}_n,\bm{x}_{n-1},\ldots,\bm{x}_1)}
\nonumber\\
&\leq&
\left[
\alpha_h
\left(
\gamma^n \|\bm{y}_0\| + \alpha_{\sigma}\,\sum_{i=1}^n \,\gamma^i \|\bm{x}_i\|
\right)
+\kappa\right]^2.
\label{eq:fbound2}
\eeqa
Now, in view of Kolmogorov two-series theorem~\cite{FellerBookV2}, it makes sense to introduce the random variable:
\beq
\bm{\xi}\dfz\sum_{i=1}^\infty \,\gamma^i \|\bm{x}_i\|,
\eeq 
which has the first two moments bounded since $\gamma<1$ and $\E[\|\bm{x}\|^2]<\infty$. 
Therefore, from~(\ref{eq:fbound2}) we have:
\beq
\bm{u}'_n\leq 
\left[
\alpha_h
(
\|\bm{y}_0\| + \alpha_{\sigma}\,\bm{\xi}
)
+\kappa\right]^2
\eeq
This clearly shows that:
\beq
\bm{u}'_n\leq \bm{u}',
\eeq 
where $\bm{u}'$ is an integrable random variable that does not depend on $n$. 
This implies that $\bm{u}'_n$ are uniformly integrable. Since uniform integrability is a property of the distribution, also the original $\bm{u}_n$ are uniformly integrable, and the claim of the lemma follows.
\end{IEEEproof}

\begin{IEEEproof}[Proof of Proposition~\ref{theor:integrability}]
We need to show that the following expectations are well-defined:
\beq
\mathcal{F}_0(n)=\E[h(\bm{y}_n) h(\bm{y}_n)^{\top}],~~
\E[\omega(\bm{y}_n) h(\bm{y}_n)^{\top}].
\eeq
That $\mathcal{F}_0(n)$ is well-posed follows directly from Lemma~\ref{lem:UIh2}. 
For what concerns $\E[\omega(\bm{y}_n) h(\bm{y}_n)^{\top}]$, we know that $\|\omega(\bm{y}_n)\|^2$ is integrable in view of Lemma~\ref{lem:oneoverh1}, while $\|h(\bm{y}_n)\|^2$ is integrable in view of Lemma~\ref{lem:UIh2}. 
Thus, the claim follows since the product of two $L_2$ functions is $L_1$. 
In a more explicit form, if we take the $(i,j)$-th entry of the matrix $\omega(\bm{y}_n) h(\bm{y}_n)^{\top}$, we get:
\beq
w_i(\bm{y}_i(n)) h_j(\bm{y}_j(n)), 
\eeq
and:
\beq
|\E[w_i(\bm{y}_i(n)) h_j(\bm{y}_j(n))]|
\leq
\sqrt{\E[w^2_i(\bm{y}_i(n))] \,\E[h^2_j(\bm{y}_j(n))]}
\eeq
by simple application of the Cauchy-Schwartz inequality.
\end{IEEEproof}

\section{Proof of Proposition~\ref{theor:invertibility}}
\label{app:invertibility}

\begin{IEEEproof}[Proof of Proposition~\ref{theor:invertibility}]
We will prove the claim by contradiction. 
Assume $\mathcal{F}_0(n)$ not invertible. In view of~(\ref{eq:vorthogas}), this corresponds to saying that
$h(\bm{y}_n)^{\top}v=0$ almost surely, i.e.:
\beq
\P[h(\bm{y}_n)\in \mathcal{V}^{\perp}]=1,
\label{eq:H0empty}
\eeq
where:
\beq
\mathcal{V}^{\perp}=\{z: z^{\top}v=0\}.
\eeq
On the other hand, we have that ($\mu_{\textnormal{Leb}}$ is the Lebesgue measure in $\mathbb{R}^N$):
\beq
\begin{array}{ccc}
&\mu_{\textnormal{Leb}}(\mathcal{V}^{\perp})=0&\\
&\Downarrow&\\ 
&\mu_{\textnormal{Leb}}(h^{-1}(\mathcal{V}^{\perp}))=0&\\
&\Downarrow&\\ 
&\P[\bm{y}_n\in h^{-1}(\mathcal{V}^{\perp})]=0,&
\end{array}
\eeq
where the first equality holds true because $\mathcal{V}^{\perp}$ is a lower-dimensional subspace of $\mathbb{R}^N$; the intermediate implication comes from Assumption~\ref{assum:assumh2}; whereas the last implication comes from the fact that $\bm{y}_n$ is absolutely continuous with respect to the Lebesgue measure, as we will now show. 
Indeed, from~(\ref{eq:ultimodelvec}) we can write:
\beq
\bm{y}_{n} = \sigma\left( g(\bm{y}_{n-1}) \odot A h(\bm{y}_{n-1}) +\bm{x}_{n}\right).
\eeq
Now, the argument of the function $\sigma$ is absolutely continuous because it is the sum of two independent random variables, with one of these being absolutely continuous. Since the mapping $\sigma$ is a diffeomorphism, it preserves absolute continuity. Indeed, letting $\bm{y}=\sigma(\bm{z})$, with $\bm{z}$ being absolutely continuous with a density $f_z$, we have that (${\sf D}$ denotes the Jacobian): 
\beqa
\P[\bm{y}\in \mathcal{A}] & = & \P[\sigma(\bm{z})\in \mathcal{A}]=\P[\bm{z}\in \sigma^{-1}(\mathcal{A})]\nonumber\\
&=&\int_{\sigma^{-1}(\mathcal{A})} f_z(\zeta) d\zeta\nonumber\\
&=&\int_{\mathcal{A}} f_z(\sigma^{-1}(y)) \left|\det({\sf D}\sigma^{-1}(y))\right|dy,
\label{eq:JacobianTransform}
\eeqa
and absolute continuity of $\bm{y}=\sigma(\bm{z})$ follows by absolute continuity of $\bm{z}$. 
We conclude that $\bm{y}_n$ is absolutely continuous. 
But we have the equality:
\beq
\P[\bm{y}_n\in h^{-1}(\mathcal{V}^{\perp})]=
\P[h(\bm{y}_n)\in \mathcal{V}^{\perp}],
\eeq
which violates condition~(\ref{eq:H0empty}), yielding a contradiction.
\end{IEEEproof}

\section{Proof of Proposition~\ref{theor:ergodicity}}
\label{app:ergodicity}
\begin{IEEEproof}
In order to prove ergodicity, it is convenient to use the additive noise model representation in~(\ref{eq:additivenoisemodel}), which we report here for convenience:
\beq
\bm{z}_{n+1}=G(\bm{z}_n) + \bm{x}_{n+1},
\label{eq:anm2}
\eeq
where we have used the definitions, 
\beq
\bm{z}_n=\sigma^{-1}(\bm{y}_n),~~\widetilde{g}(y)=g(\sigma(y)), ~~ \widetilde{h}(y)=h(\sigma(y)),
\eeq  
and, for $z\in\mathbb{R}^N$:
\beq
G(z)=\widetilde{g}(z) \odot A \widetilde{h}(z).
\eeq
In view of Example 7.4.6 in~\cite{FurtherMarkovBook}, if $F$ is continuous, if the noise $\bm{x}_n$ is absolutely continuous with respect to the Lebesgue measure with almost everywhere positive density with finite mean, and if  there exist positive constants $\beta$ and $\gamma<1$ such that\footnote{Actually, condition~(\ref{eq:suffcondstab}) rephrases the condition reported in~\cite{FurtherMarkovBook} in a way that is more convenient in our setting.}: 
\beq
\E[\|G(z)+\bm{x}\|]\leq \gamma\|z\| +\beta,
\label{eq:suffcondstab}
\eeq
then the chain is $w$-geometrically ergodic with weight function $\omega(z)=1+|z|$~\cite{FurtherMarkovBook}. 
Since by assumption continuity of $F$ and the properties of the noise are fulfilled, it remains to verify that~(\ref{eq:suffcondstab}) holds true in our setting. Reasoning as done in~(\ref{eq:chaineq}), we get:
\beqa
\lefteqn{\|G(z)+\bm{x}\|\leq\| D_{g}(\sigma(z)) A h(\sigma(z)) \|+\|\bm{x}\|} 
\nonumber\\
&\leq& 
\|A\|\, \| D_{g}(\sigma(z))\| \, \|h(\sigma(z))\| +\|\bm{x}\|
\nonumber\\
&\leq&
\|A\|\, (\alpha_h\| \sigma(z)\|^p + \beta_h) \, (\alpha_g\| \sigma(z)\|^q + \beta_g)+\|\bm{x}\|
\nonumber\\
&\leq&
\|A\|\alpha_h\alpha_g \|\sigma(z)\| + 
\|A\|\beta_h\beta_g+\|\bm{x}\|
\nonumber\\
&+&
\|A\|\alpha_h\beta_g\|\sigma(z)\|^p
+
\|A\|\alpha_g\beta_h\|\sigma(z)\|^q.
\label{eq:chaineq2}
\eeqa
We shall consider the case $p>0$, $q>0$. The proof for the remaining cases is similar. 
Reasoning along the same lines as in~(\ref{eq:boundingpq}), we can conclude that, for an arbitrarily small $\epsilon>0$:
\beq
\|G(z)+\bm{x}\| \leq (\|A\|\alpha_h\alpha_g + \epsilon) \|\sigma(z)\| + c+\|\bm{x}\|,
\eeq
where $c$ is a certain positive constant. 
On the other hand, in view of~(\ref{eq:newLipschitzlikesigma})--(\ref{eq:newLipschitzlikeh0}) we can further write:
\beq
\E[\|G(z)+\bm{x}\|] \leq \gamma \|z\|+ \beta,
\eeq
where $\gamma=\|A\|\alpha_h\alpha_g\alpha_\sigma + \epsilon'$, and $\beta$ is a proper constant. Therefore, we have proved that~(\ref{eq:suffcondstab}) is verified, which implies $w$-ergodicity of $\bm{z}_n$ in light of Example 7.4.6 in~\cite{FurtherMarkovBook}. Since $\bm{z}_n=\sigma^{-1}(\bm{y}_n)$, we have in fact proved $w$-ergodicity for $\bm{y}_n$.

Since we now have $w$-geometric ergodicity, we also have convergence in total variation to the unique invariant measure $\pi_y$, i.e.,
\beq
\|\nu P^{n}-\pi_y \|_{{\sf TV}}\overset{n\rightarrow \infty}\longrightarrow 0
\label{eq:mainTV}
\eeq
for any initial distribution $\nu$, where $P$ is the transition kernel characterizing the Markov process $\left\{\bm{y}_n\right\}$, and $\|\cdot\|_{{\sf TV}}$ is the total variation norm~\cite{FurtherMarkovBook}. 
In other words, the unique invariant measure $\pi_y$ is a {\em global attractor} of the dynamical system $\nu P^n$ in the space of probability measures on $\mathbb{R}^N$.

Now, the aforementioned convergence in total variation will imply the strong law in~(\ref{eq:asconvtheorergF0}) via application of the ergodic theorem~\cite{MeynTweedie, MCandInvProb}, if we prove that the following moment (the notation $\mathbb{E}_{\pi_y}[\cdot]$ means that the expectation of the random variable under brackets is computed under the measure $\pi_y$):
\begin{equation}
\mathcal{F}_0\dfz\mathbb{E}_{\pi_y}\left[h(\bm{y})h(\bm{y})^{\top}\right]
\label{eq:F0pinvariant}
\end{equation}
is well-defined. 
To this end, let us first observe that the constant $\kappa$ appearing in~(\ref{eq:kappadef}) is strictly smaller than $1$ by assumption, which implies that, for a sufficiently small $\epsilon$, it is possible to choose a constant $\gamma<1$ in Lemma~\ref{lem:UIh2}. 
Therefore, we have that the sequence of random variables $\left|\left|h(\bm{y}_n)\right|\right|^2$ is uniformly integrable.
Now, since: $i)$ $\bm{y}_n$ converges in total variation (and, hence, in distribution); $ii)$ $h$ is a continuous mapping; $iii)$ the sequence $\left|\left|h(\bm{y}_n)\right|\right|^2$ is uniformly integrable, we can conclude that the expectation in~(\ref{eq:F0pinvariant}) is well-defined, and that the first convergence of expectations in~(\ref{eq:F0F1expeconv}) holds true~\cite{DasGupta2008}.
We switch to the analysis of the one-lag matrix function. First, in view of~(\ref{eq:ggrangerdirect}), we have:
\beq
\mathcal{F}_1(n)=A \mathcal{F}_0(n)\rightarrow A \mathcal{F}_0,
\eeq
since we have proved that $\mathcal{F}_0(n)\rightarrow \mathcal{F}_0$. 
Second, we want to show that~(\ref{eq:asconvtheorergF1}) holds true. 
By applying the definitions in~(\ref{eq:zerolag0}) and~(\ref{eq:onelag0}) to~(\ref{eq:identity2}), we get:
\beq\label{eq:identity3}
F_1(\bm{y}_{n+1},\bm{y}_n)
=
A F_0(\bm{y}_n)\mathbb{I}[\bm{y}_n\notin\mathcal{Z}] + e(\bm{x}_{n+1},\bm{y}_n),
\eeq
where we have defined the quantity:
\beq
e(\bm{x}_{n+1},\bm{y}_n)=\bm{x}_{n+1}\odot \omega(\bm{y}_n) h(\bm{y}_n)^{\top}\mathbb{I}[\bm{y}_n\notin\mathcal{Z}].
\label{eq:errdef}
\eeq
In light of~(\ref{eq:empirF0F1}), from~(\ref{eq:identity3}) we can also write:
\beq
\widehat{\mathcal{F}}_1(n)=A\,\displaystyle{\frac{\sum_{k=0}^{n-1}F_0(\bm{y}_k)\mathbb{I}[\bm{y}_k\notin\mathcal{Z}]}{n}}
+\displaystyle{\frac{\sum_{k=0}^{n-1} e(\bm{x}_{k+1},\bm{y}_k)}{n}}.
\label{eq:newdecomposition}
\eeq
Since uniform integrability is not impaired by the presence of the indicator, and since the probability that $\bm{y}_k\notin\mathcal{Z}$ is equal to one, from the previous analysis it is clear that:
\beq
\displaystyle{\frac{\sum_{k=0}^{n-1}F_0(\bm{y}_k)\mathbb{I}[\bm{y}_k\notin\mathcal{Z}]}{n}}\stackrel{\textnormal{a.s.}}{\longrightarrow} A\mathcal{F}_0.
\eeq 
Therefore, we need to establish that the second term on the RHS in~(\ref{eq:newdecomposition}) converges to zero almost surely.
To this aim, let us consider the joint process $\bm{v}_n\in\mathbb{R}^{2 N}$: 
\beq
\bm{v}_n\dfz(\bm{y}_n,\bm{x}_{n+1}).
\eeq 
Since $\bm{y}_n$ and $\bm{x}_{n+1}$ are statistically independent, and since the vectors $\bm{x}_{n+1}$ are i.i.d. across time (with a certain measure $\pi_x$), it is immediately seen that $\bm{v}_n$ admits a unique invariant measure, given by the product measure $\pi_v=\pi_x\times\pi_y$, between the invariant measure of $\bm{y}_n$ and the (stationary) measure of $\bm{x}_{n+1}$.
Moreover, for any $n$ we obvisouly have that $\pi_{v_n}=\pi_{x}\times \pi_{y_n}$, where $\pi_{y_n}=\nu P^n$ is the distribution of $\bm{y}_n$, for a certain initial distribution $\nu$. 
Using then~(\ref{eq:mainTV}), we can conclude that:
\beq
\|\pi_{v_n}-\pi_{v} \|_{{\sf TV}}\overset{n\rightarrow \infty}\longrightarrow 0.
\eeq
Therefore, we can apply the ergodic theorem~\cite{MeynTweedie, MCandInvProb} to $\bm{v}_n$. In particular, we can observe that:
\beq
\frac{1}{n}\sum_{k=0}^{n-1} e(\bm{y}_k,\bm{x}_{k+1})\stackrel{\textnormal{a.s.}}{\longrightarrow} \E_{\pi_x\times\pi_y}[e(\bm{x},\bm{y})],
\label{eq:ergoF1new}
\eeq
provided that the latter expectation, computed under the invariant distribution, exists. 
However, from~(\ref{eq:errdef}) we can write:
\beq
\E[e(\bm{x}_{n+1},\bm{y}_n)]=\E[{\sf diag}(\bm{x}_{n+1})]\E[\omega(\bm{y}_n)h(\bm{y}_n)^{\top}],
\eeq
since integrability of the pertinent random variables has been already established in Lemma~\ref{theor:GG}. 
Moreover, since we have also established that $\omega(\bm{y}_n)$ and $h(\bm{y}_n)$ are $L_2$-integrable, with integrals that are bounded by a constant independent of $n$, joint application of Skorohod's representation theorem and by Fatou's lemma~\cite{DasGupta2008} imply that also the expectation $\E_{\pi_y}[\omega(\bm{y})h(\bm{y})^{\top}]$ (i.e., computed under the invariant distribution $\pi_y$) exists. 
But since the invariant distribution $\pi_v$ has the product form $\pi_x\times\pi_y$, we can write:
\beq
\E_{\pi_x\times\pi_y}[e(\bm{x},\bm{y})]=\underbrace{\E_{\pi_x}[{\sf diag}(\bm{x})]}_{=0}\E_{\pi_y}[\omega(\bm{y})h(\bm{y})^{\top}]=0,
\eeq
which applied in~(\ref{eq:ergoF1new}) yields the desired claim.

In order to conclude the proof of the proposition, we have to prove invertibility of $\mathcal{F}_0$. 
To this aim, let us consider an initial state $\bm{y}_0$ distributed according to the invariant measure $\pi_y$. In view of~(\ref{eq:ultimodelvec}), we have:
\beq
\bm{y}_{1} = \sigma\left( g(\bm{y}_{0}) \odot A h(\bm{y}_{0}) +\bm{x}_{1}\right),
\label{eq:y1form}
\eeq
and we observe that the state $\bm{y}_1$ will be still distributed according to $\pi_y$ due to invariance, implying that we can write 
\beq
\mathcal{F}_0=\mathbb{E}_{\pi_y}\left[h(\bm{y})h(\bm{y})^{\top}\right]=\mathbb{E}\left[h(\bm{y}_1)h(\bm{y}_1)^{\top}\right].
\eeq
Therefore, the fact that $\mathcal{F}_0$ is invertible follows by applying to the random variable in~(\ref{eq:y1form}) the reasoning used to prove Proposition~\ref{theor:invertibility}.
\end{IEEEproof}


\end{document}